\documentclass[5p,times,authoryear]{elsarticle}

\usepackage[english]{babel}     
\addto\captionsenglish{%
}
\usepackage[latin1]{inputenc}   
\usepackage{amsmath}            
\usepackage{epstopdf}           
\usepackage{flushend}           
\usepackage{hyperref} 
\usepackage{multicol, blindtext}

\hypersetup{
    colorlinks=true,
    linkcolor=blue,
    filecolor=magenta,      
    urlcolor=cyan,
}

\usepackage{dblfloatfix}    

\usepackage{amssymb}

\usepackage[figuresright]{rotating}

\usepackage{subfigure}

\begin{document}

\begin{frontmatter}






\title{Ares: A Mars model  retrieval  framework for ExoMars Trace Gas Orbiter NOMAD solar occultation measurements}


\author[First,Second]{George Cann\corref{cor1}}
\ead{george.cann.15@ucl.ac.uk}

\author[Second]{Ahmed Al-Refaie}
\ead{ahmed.al-refaie.12@ucl.ac.uk}

\author[Second]{Ingo Waldmann}
\ead{ingo.waldmann@ucl.ac.uk}

\author[First]{Dave Walton}
\ead{d.walton@ucl.ac.uk}

\author[First]{Jan-Peter Muller}
\ead{j.muller@ucl.ac.uk}

\address[First]{Imaging Group, Mullard Space Science Laboratory, Department of Space and Climate Physics, University College London, Holmbury St. Mary, Surrey, RH5 6NT, UK. Email: george.cann.15@ucl.ac.uk.}

\address[Second]{Astrophysics Group, Department of Physics and Astronomy, University College London, Gower Street, WC1E 6BT, UK.}

\begin{abstract}
Ares is an extension of the TauREx 3 retrieval framework for the Martian atmosphere. Ares is  a collection of new atmospheric parameters and forward models, designed for the European Space Agency's (ESA) Trace Gas Orbiter (TGO) Nadir and Occultation for MArs Discovery (NOMAD) instrument, Solar Occultation (SO) channel. Ares provides unique insights into the chemical composition of the Martian atmosphere by applying methods utilised in exoplanetary atmospheric retrievals, \cite{Wal:15}, \cite{Al-R:19}. This insight may help unravel the true nature of $\textup{CH}_{4}$ on Mars. The Ares model is here described. Subsequently, the components of Ares are defined, including; the NOMAD SO channel instrument function model, Martian atmospheric molecular absorption cross-sections, geometry models, and a NOMAD noise model. Ares atmospheric priors and forward models are presented, (i.e., simulated NOMAD observations), and are analysed, compared and validated against the Planetary Spectrum Generator,  \cite{Vil:18}.
\end{abstract}

\begin{keyword}

Mars, $\textup{CH}_{4}$, Ares, TauREx, NOMAD.
\end{keyword}

\end{frontmatter}

\section{Introduction}

In 2003, methane, $\textup{CH}_{4}$, was tentatively detected in the Martian
atmosphere, at 10 ppbv $\pm$ 5, varying by up to 30 ppbv globally, \cite{For:04}, and 10 ppbv $\pm$ 3 \cite{Kras:04}. $\textup{CH}_{4}$ has, at most, a predicted photochemical lifetime of 300  years according to \cite{Sum:02}. This implies that  $\textup{CH}_{4}$ in the Martian atmosphere  should be uniformly distributed over Mars. However, non-uniform distributions of $\textup{CH}_{4}$ have been observed, \cite{Mum:09}. This raises questions with regard to the source(s) and/or sink(s) of $\textup{CH}_{4}$. Abiotic and biotic sources have been suggested to explain the detection, ranging from olivine serpentinization, \cite{Mor:07}, to methanogenesis by methanogenic archaea, \cite{All:06}.

The Nadir and Occultation for MArs Discovery (NOMAD) instrument, \cite{Van:15}, \cite{Rob:16}, onboard the European Space Agency's Exomars Trace Gas Orbiter (TGO) was designed to investigate the nature of methane, $\textup{CH}_{4}$, on Mars, \cite{Liu:19}, \cite{Mum:09}. However, the arrival of TGO and subsequent science mission has detected no $\textup{CH}_{4}$, with an upper limit of 0.05 ppbv, \cite{Kor:19}. In contrast, NASA's Curiosity Sample Analysis at Mars Tunable Laser Spectrometer instrument (SAM-TLS), \cite{Mah:12}, has made multiple measurements of $\textup{CH}_{4}$, including measuring an elevated $\textup{CH}_{4}$ background of 7.2 $\pm$ 2.1 ppbv $\textup{CH}_{4}$ over a 60-sol period in 2013, \cite{Web:15}. Subsequently, using SAM-TLS, \cite{Web:18}, determined a mean $\textup{CH}_{4}$ abundance of 0.41 $\pm$ 0.16 ppbv, as well as a repeatable seasonal variation from 0.24 to 0.65 ppbv. Moreover, on $\textup{19}^{\textup{th}}$ June 2019 it was reported that SAM-TLS measured a spike of 21 ppbv at Teal Ridge in Gale Crater, \cite{NASA:19}. Furthermore, the Planetary Fourier Spectrometer (PFS), \cite{For:97}, onboard Mars Express measured 15.5 $\pm$ 2.5 ppbv of $\textup{CH}_{4}$, above Gale Crater on $\textup{16}^{\textup{th}}$ June 2013, one day after SAM-TLS independently detected a $\textup{CH}_{4}$ spike of 5.78 $\pm$ 2.27 ppbv, \cite{Giu:19}. Since then PFS has detected no $\textup{CH}_{4}$, \cite{ESA:19}. The discrepancy between surface measurements by SAM-TLS and orbital measurements from NOMAD and PFS, combined with the independent confirmation of detection of $\textup{CH}_{4}$ by PFS, significantly constrains the mechanisms to corroborate the measurements. 

Here we present a new retrieval scheme called Ares, an extension to the TauREx framework designed for TGO NOMAD Solar Occultation (SO) channel solar occultation measurements. Ares allows atmospheric sounding of the Martian atmosphere by applying methods developed originally for extracting tiny signals from noisy measurements of in exoplanetary atmospheric retrievals. This insight could help unravel the nature of $\textup{CH}_{4}$ on Mars. 

This paper is organized into distinct sections; firstly, the Ares model is described; subsequently, the components of Ares and associated classes are defined, including; the NOMAD instrument function model, HITRAN 2016, \cite{Gor:16}, Martian absorption cross-sections, the geometry modules. Secondly, simulated NOMAD observations with Ares, are analysed, compared, and validated against the Planetary Spectrum Generator, \cite{Vil:18}. 

\section{Ares outline}

 Ares is an extension of TauREx3, \cite{Al-R:19}, the $\textup{3}^{\textup{rd}}$ generation of TauREx. TauREx, (Tau Retrieval for Exoplanets), \cite{Wal:15}, is a fully Bayesian atmospheric retrieval framework that uses Nested Sampling, \cite{Ski:06, Fer:09}, and Markov chain Monte Carlo (MCMC) methods to sample the full likelihood space of possible solutions. This allows TauREx to produce marginalised and conditional posterior distributions of forward model parameters, which can be used to map correlations between forward model parameters. This is advantageous over other planetary atmospheric retrieval frameworks that only find the \textit{maximum a posteriori} (MAP) solution through Optimal Estimation, \cite{Rod:00}. The work presented here focuses on extending the TauREx3 radiative transfer forward models for Mars retrievals. We have named this TauREx\,3 Mars extension module Ares.
 
 \begin{figure}[ht]
\centering
  \includegraphics[width=8.9cm]{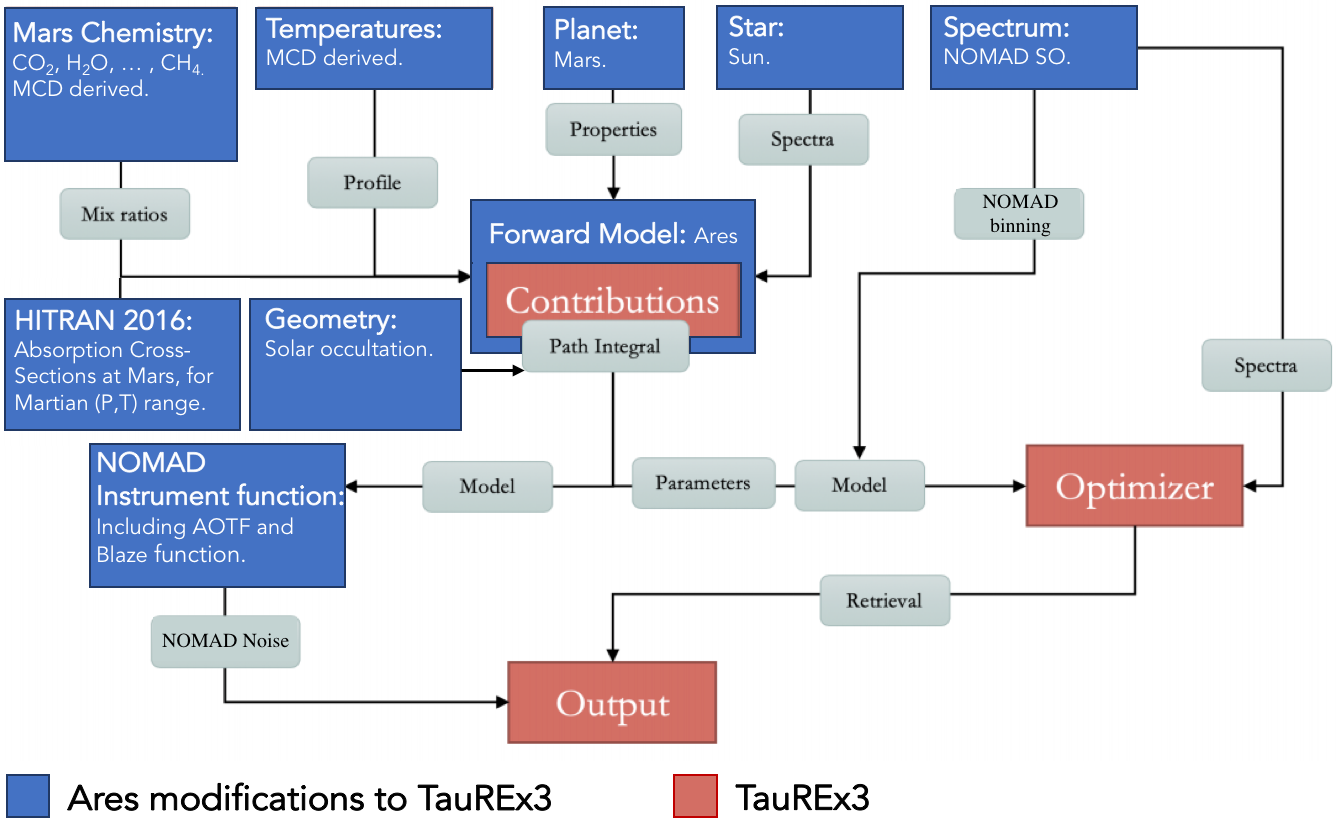}\\
  \caption{A flow diagram of Ares, modified from \cite{Al-R:19}. Red boxes represent TauREx3 modules. Blue boxes represent Ares modifications to TauREx3.}\label{fig1}
\end{figure}
 
 The NOMAD SO channel blaze function, acousto-optic tunable filter (AOTF) modules have been incorporated into the Ares  forward model. Furthermore, Martian ellipsoidal and spherical geometry modules have been included in order to calculate the atmospheric layer line-of-sight (LOS) intersection points and corresponding path lengths, \cite{Tho:18}. A Mars Climate Database (MCD) Python library, \texttt{pymcd}, has been written and linked to the RTM, providing access to Martian vertical temperature, pressure and volume-mixing-ratio (VMR) profiles, \cite{Forg:99}, \cite{Mill:18}. Moreover, Martian absorption cross-sections have been generated using HITRAN 2016, \cite{Gor:16}, molecular line lists and ExoCross, \cite{Yur:18}, accounting for expected Martian pressures, temperatures and $\textup{CO}_{2}$ broadening, due to the $\textup{CO}_{2}$-rich Martian atmosphere. Overall this enables Ares to produce simulated NOMAD SO channel transmission spectra.

\subsection{NOMAD instrument}

NOMAD is a high-resolution echelle grating spectrometer suite consisting of three channels; namely the Ultraviolet and Visible Spectrometer (UVIS), Solar Occultation (SO) and Limb, Nadir and Occultation (LNO), \cite{Rob:16}. Each channel can operate in different geometry modes. UVIS functions in nadir and solar occultation mode geometries from 0.2-0.65 $\mu$m, SO functions in solar occultation geometry from 2.3-4.3 $\mu$m and LNO is dedicated to operating in limb and nadir geometries also from 2.3-4.3 $\mu$m. NOMAD is dedicated to investigate the composition of trace gas species in the Martian atmosphere, which can provide  insights into present day geological and biological processes. NOMAD targets trace gas species including  molecules such as, CO, HDO, $\textup{C}_{2}\textup{H}_{2}$, $\textup{C}_{2}\textup{H}_{4}$, $\textup{H}_{2}\textup{CO}$, $\textup{H}_{2}\textup{S}$, $\textup{HCl}$, HCN, $\textup{HO}_{2}$, $\textup{NH}_{3}$, $\textup{N}_{2}\textup{O}$, $\textup{NO}_{2}$, OCS and $\textup{O}_{3}$. Additionally, NOMAD should be capable of determining the abundances of the isotopologues of the aforementioned species, that are of particular relevance to determining the biotic or abiotic nature of $\textup{CH}_{4}$, \cite{Van:15}, \cite{Rob:16}. These include the isotopologues of the Martian atmospheric carbon sources $\textup{CO}_{2}$, $\textup{CO}$, and $\textup{CH}_{4}$, namely $^{13}\textup{CO}_{2}$, $^{17}\textup{OCO}$, $^{18}\textup{OCO}$, $^{18}\textup{CO}_{2}$, $^{13}\textup{CO}$,  $^{18}\textup{CO}$, $^{12}\textup{CH}_{4}$, $^{13}\textup{CH}_{4}$ and $\textup{CH}_{3}\textup{D}$. Furthermore, NOMAD should be capable of determining the abundances of hydrocarbons such as $\textup{C}_{2}\textup{H}_{6}$ and sulfur sources, such as  $\textup{SO}_{2}$, \cite{Yun:18}.

\subsubsection{SO channel}

This study will predominately focus on the NOMAD SO channel. NOMAD's SO channel covers 2.3-4.3 $\mu$m, for diffraction orders 96-225, with a resolving power of $\approx$ 20,000 and a radio frequency input to the SO AOTF ranging between 12,300-31,100 kHz. For comparison, the LNO channel covers diffraction orders 108-220, with a resolving power of $\approx$ 10,000 and a radio frequency input to the LNO AOTF ranging between 14,200-32,100 kHz. The NOMAD SO channel detector consists of a grid of 320 columns and 256 rows, along the spectral and spatial dimensions, respectively.

For NOMAD's SO channel, typically 24 rows of the detector range over an interval of approximately 7.5\,km along the tangent height path, with a vertical sampling of 500\,m, with the number of rows illuminated dependent on the Mars-Sun distance. Whilst extensive definitions of the instrument are given in \cite{Nef:15}, \cite{Rob:16} and \cite{Liu:19}, the temperature dependence of the AOTF and blaze functions are not reported. We obtained the requisite details on how to define the AOTF and blaze functions accounting for  their temperature dependencies from the NOMAD Experiment to Archive Interface Document (EAICD), \cite{Tho:18}, and the NOMAD Datasets and Calibration Steps document, \cite{Tho:19}. 

\subsubsection{Calibration}

NOMAD SO channel calibrations have been performed by, \cite{Liu:19}, \cite{Tho:18}, and include the conversion of detector pixel number and diffraction order to wavenumber, i.e. instrument spectral calibration, as well as instrument spectral resolution determination and thermal effects quantification. Moreover, measurements of the tuning relation have been performed to determine the relationship between the radio frequency applied to the SO channel AOTF and wavenumber at which the SO channel AOTF transfer function peaks. 

The main product derived from the NOMAD instrument calibrations is an accurate model for NOMAD SO channel spectra. \cite{Liu:19} and \cite{Tho:18} note that different diffraction orders mix to produce the observed spectra, where between 15 - 50 \% in a target order is contributed to by nearby orders, a property that increases with order number and AOTF frequency. \cite{Liu:19} gives the functional relationship between pixel number $p$ (from 0 to 319) wavenumber $\nu$ and order number $m$ being modelled by a $2^{\textup{nd}}$ order polynomial. The retrieved coefficients for the aforementioned spectral calibration are given by,

\begin{equation}
\label{equ:spec_cal}
    \frac{\nu }{m} = F_{0} + F_{1}p + F_{2}p^{^{2}}.
\end{equation}

Similarly the tuning relation is given by the following $2^{\textup{nd}}$ order polynomial,

\begin{equation}
    \nu = G_{0} + G_{1}A + G_{2}A^{^{2}},
\end{equation}

where $A$ is the AOTF frequency.  The retrieved coefficients, $G_{0}$, $G_{1}$, $G_{2}$, $F_{0}$, $F_{1}$ and $F_{2}$ for the spectral calibration can be found in Table 1. and 2. of \cite{Liu:19}. Spectral calibration of NOMAD's SO channel is challenging due to the presence of a blaze function, an AOTF and the effects of temperature variation on these functions.

\subsubsection{AOTF Transfer Function}

NOMAD's SO channel AOTF is fundamentally a filter, as without it associating a particular absorption line to a particular wavelength would be challenging, as more than 100 diffraction orders would simultaneously fall on the detector, \cite{Liu:19}. The AOTF transfer function is given by the following equation. 

\begin{equation}
    TF(\nu, \nu_{0},w,I_{G},\rho_{G},q,n) = F_{\textup{sinc}} + F_{\textup{gauss}} + F_{\textup{cntmn}},
\end{equation}

\noindent where $F_{\textup{sinc}}$, $F_{\textup{gauss}}$ and $F_{\textup{cntmn}}$ are given by,

\begin{align}
     F_{\textup{sinc}}(\nu, \nu_{0},I_{0},w) &= I_{0}w^{2}\frac{\left ( \sin \frac{\pi(\nu - \nu_{0})}{w} \right )^{2}}{\pi^{2}(\nu -\nu_{0})^{2}} \\
     F_{\textup{gauss}}(\nu, \nu_{0},I_{G},\sigma_{G}) &= I_{G}\exp\left ( \frac{-(\nu - \nu_{0})^{2}}{\sigma_{G}^{2}} \right ) \\
     F_{\textup{cntnm}}(\nu, \nu_{0},q,n) &= q + n(\nu -\nu_{0}),
\end{align}

\begin{figure}
\centering 
  \includegraphics[width=8.5cm]{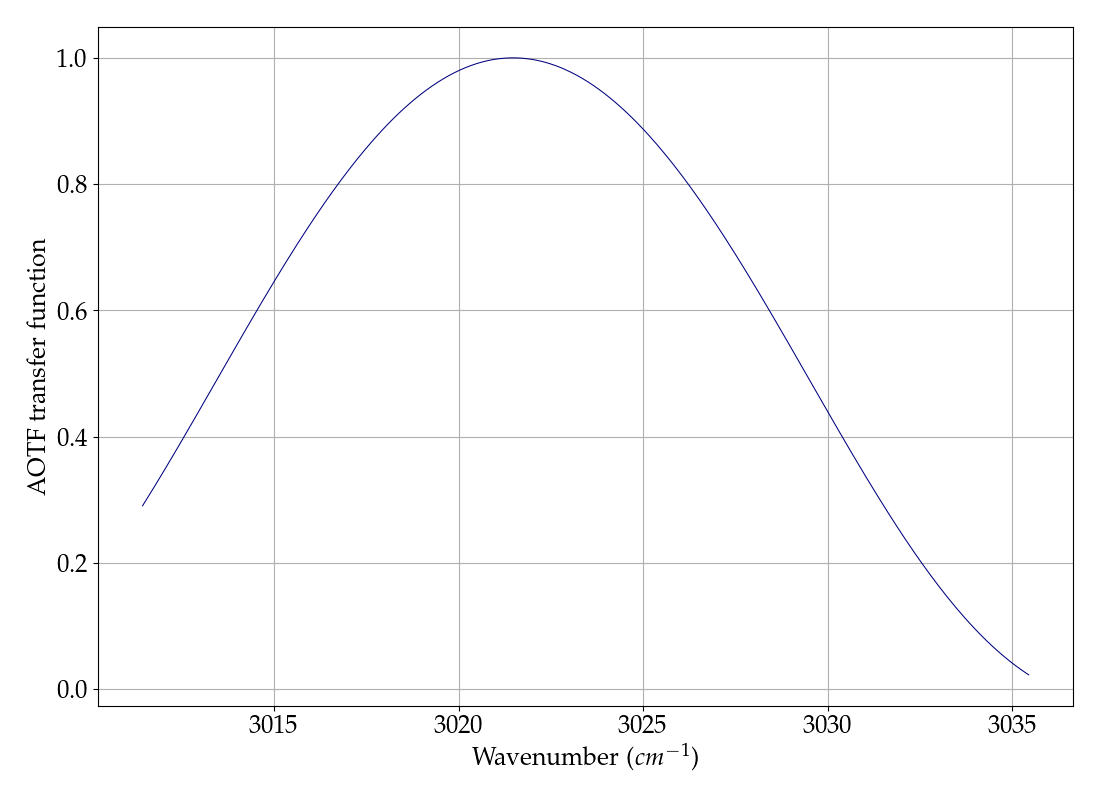}\\
  \caption{Ares derived plot of NOMAD SO channel AOTF transfer function intensity against wavenumber for diffraction order 134.}\label{fig2}
\end{figure}

\noindent where $\nu_{0}$ is the AOTF transfer function centre in $\textup{cm}^{-1}$. $I_{0}$ is the sinc-squared function amplitude. Using prior coefficients and $\nu = G_{0} + G_{1}A + G_{2}A^{^{2}}$, $\nu_{0}$ can be determined. $w$ is the location of the first zero-crossing of the sinc-squared function. $I_{G}$ is the Gaussian amplitude and $\sigma_{G}$ the Gaussian standard deviation, $q$ and $n$ are the continuum offset parameters. Finally the relationship between $w$ and the sinc-squared full-width half maximum (FWHM) is given by, $\textup{FWHM} \approx 0.886w$. The set of coefficients $G_{0}$, $G_{1}$, and $G_{2}$ are the wavenumber-AOTF frequency calibration coefficients. These terms are included in the NOMAD EAICD with the same notation and can be found in Level 2 partially processed SO HDF5 files under \texttt{WnAOTF Coefficients}.

\subsubsection{Blaze Function}
\begin{figure}
\centering
  \includegraphics[width=8.5cm]{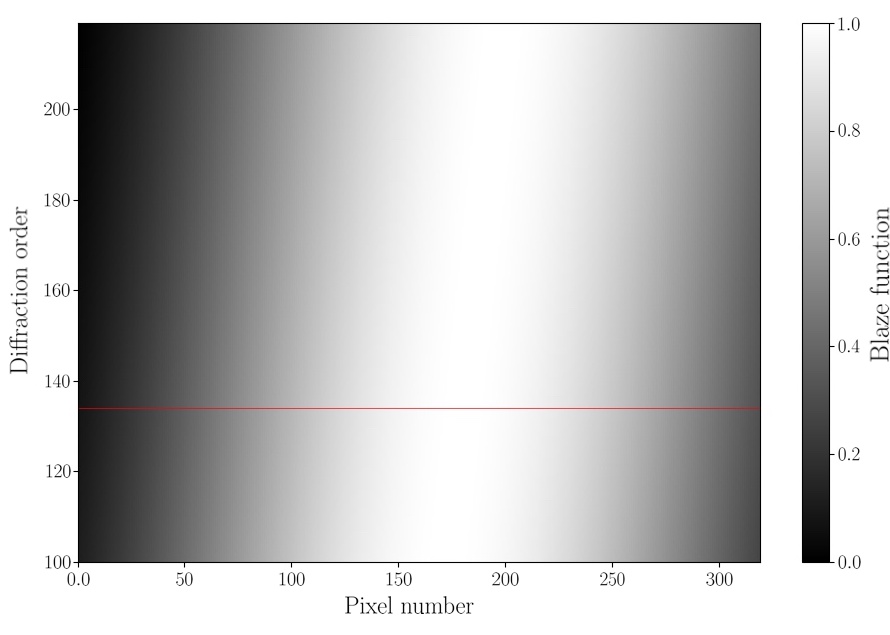}\\
  \caption{Ares derived map of NOMAD SO channel  blaze function relative intensity against diffraction order and pixel number. For comparison see Figure 11. in \cite{Liu:19}. The red line corresponds to diffraction order 134. }\label{fig3}
\end{figure}

A blaze function is included in Ares. \cite{Liu:19} give the blaze function as, 

\begin{equation}
    F_{\textup{blaze}}(p, p_{0},w_{p}) = w_{p}^{2}\frac{\left ( \sin \frac{\pi(p - p_{0})}{w_{p}} \right )^{2}}{\pi^{2}(p -p_{0})^{2}}
\end{equation}

\noindent where $p$ is the pixel number (from 0 to 319), $p_{0}$ is the centre of the function in pixel units and $w_{p}$ is the width of the blaze function. Note that in pixels $w_{p}$ is equivalent to the free spectral range. This is defined by the properties of the grating and is equivalent to $F_{0}$ in equation\,\ref{equ:spec_cal}. The blaze function $p_{0}$ is defined in the NOMAD EAICD by the following equation, 
\begin{equation}
p_{0}(m) = I_{0} + I_{1}m,
\end{equation}
where the coefficients $I_0$ and $I_1$ have been given as 150.80 and 0.22, respectively. 

\begin{figure}[h]
\centering 
  \includegraphics[width=8.8cm]{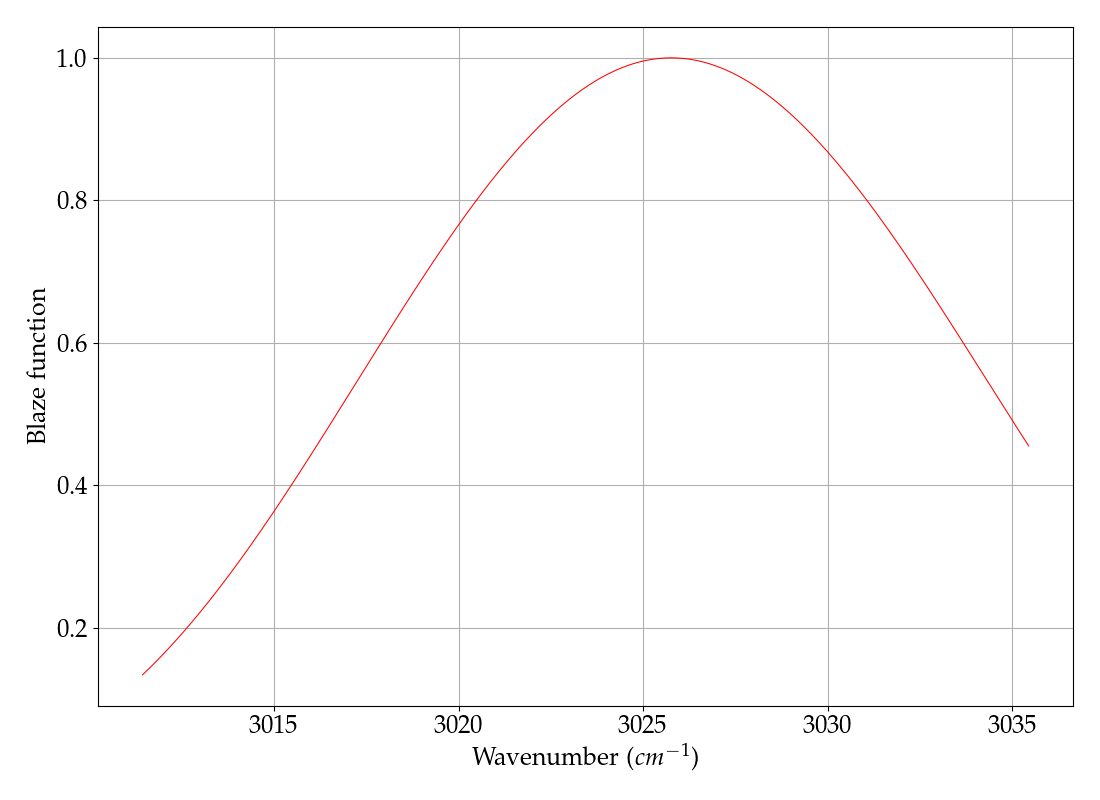}\\
  \caption{Ares derived plot of NOMAD SO channel blaze function intensity against wavenumber for diffraction order 134.}\label{fig4}
\end{figure}

\subsubsection{Spectral Continuum}

\begin{figure}
\centering
\includegraphics[width=8.8cm]{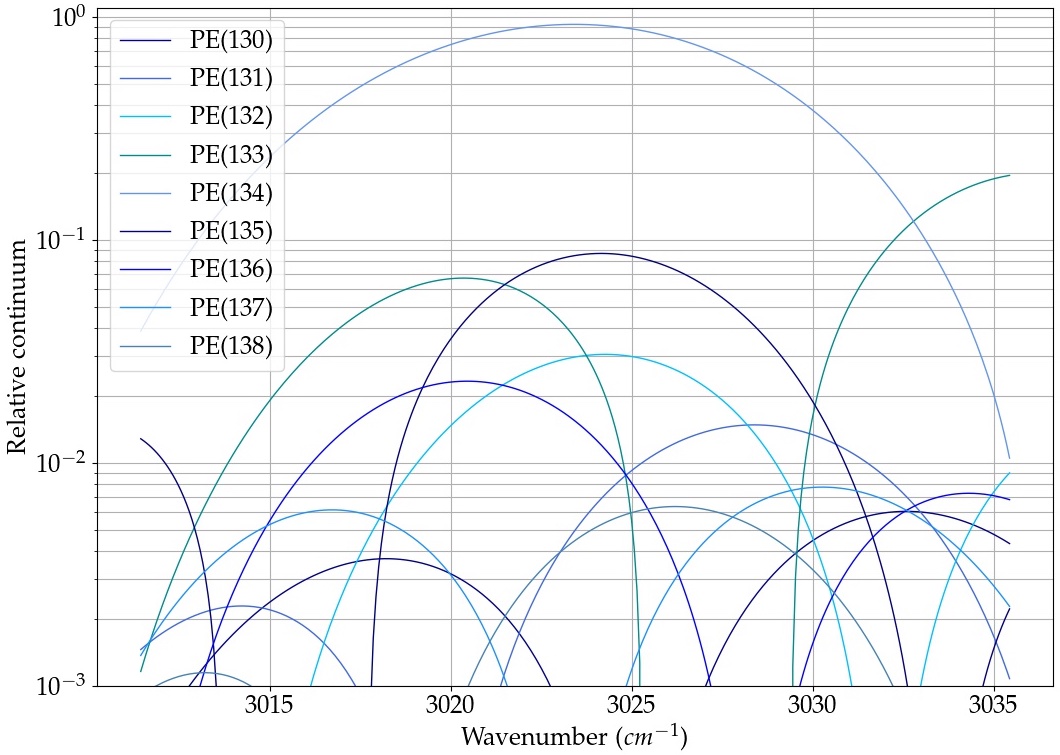}\\
\caption{Ares derived relative flux contributions, \textit{Partial Elements}, $PE(A,j)$ of the \textit{Partial Elements Continuum}, $PEC(A)$, for $j \in  [130,138]$, for diffraction order 134, with \cite{Tho:19} updated coefficients.}\label{fig5}
\end{figure}

\begin{figure}
\centering
\includegraphics[width=8.8cm]{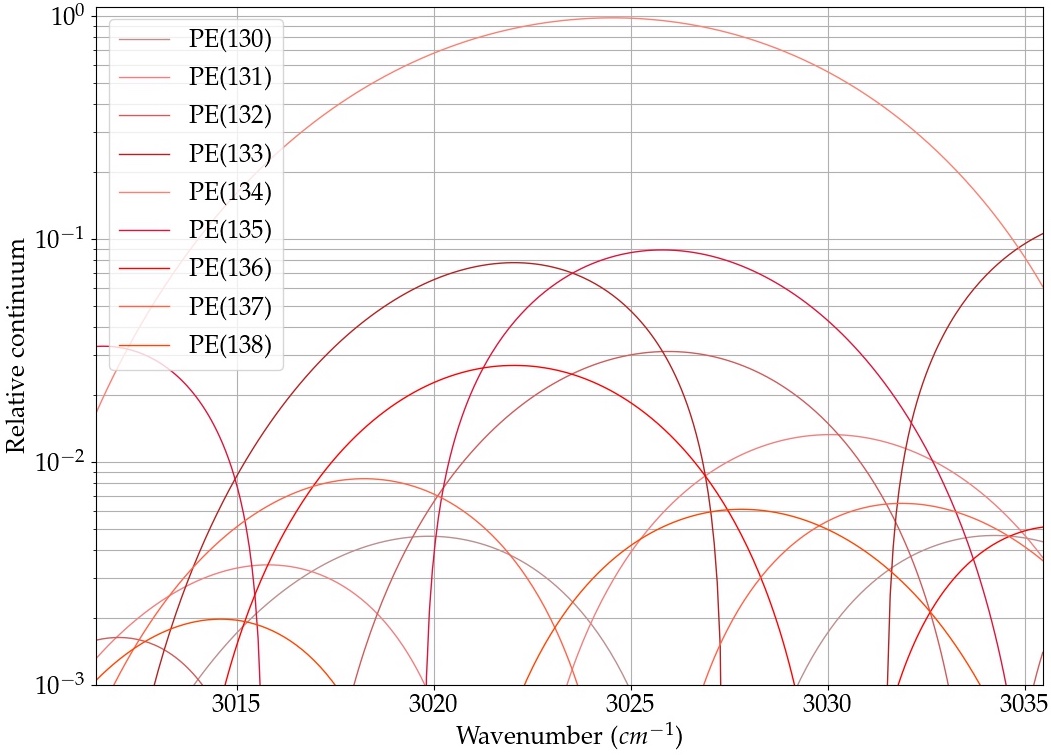}\\
\caption{Ares derived relative flux contributions, \textit{Partial Elements}, $PE(A,j)$ of the \textit{Partial Elements Continuum}, $PEC(A)$, for $j \in  [130,138]$, for diffraction order 134, with \cite{Tho:19} updated coefficients and AOTF transfer function temperature dependence.}\label{fig6}
\end{figure}

The continuum in the NOMAD SO channel spectra is the result of the incoming continuum of the Sun, modified by the SO channel AOTF transfer function and the blaze function, \cite{Liu:19}. Subsequently, the intensity of each NOMAD SO channel pixel is modulated by the intensity of the SO channel AOTF transfer function and the blaze function. The continuum in NOMAD SO channel spectra, for a particular AOTF frequency, $A$, takes the form of the \textit{Partial Elements Continuum}, \textit{PEC(A)}, \cite{Liu:19}. Where \textit{PEC(A)}, is given by, 

\begin{equation}
    PEC(A) = \sum_{j=m - \Delta m}^{m + \Delta m} PE(A,j).
\end{equation}

\begin{equation}
    PEC(A) = \sum_{j=m - \Delta m}^{m + \Delta m} AOTF (A,  \mathbf{\nu_{j}})\cdot F_{\textup{blaze}}(j,\mathbf{\nu_{j}})\cdot gain(j),
\end{equation}

\noindent where \textit{PE(A,j)} represents a \textit{Partial Element} of the \textit{PEC(A)}. The \textit{Partial Elements} for diffraction order 134 are shown in Figure \ref{fig5} and Figure \ref{fig6}. 

\subsubsection{Observed Radiance}

The radiance as observed by the NOMAD SO channel is given by \cite{Liu:19} as, 

\begin{equation}
  R(A, \mathbf{\nu_{m}}) = \sum_{j=m - \Delta m}^{m + \Delta m} AOTF (A,  \mathbf{\nu_{j}})\cdot F_{\textup{blaze}}(j,\mathbf{\nu_{j}})\cdot gain(j)\cdot R(j,\mathbf{\nu_{j}}),
\end{equation}

\noindent where $AOTF (A,  \mathbf{\nu_{j}})$ is the AOTF transfer function at the AOTF frequency $A$, for spectral grid $\mathbf{\nu_{j}}$ of diffraction order $j$. $F_{\textup{blaze}}(j,\mathbf{\nu_{j}})$ is the blaze function of diffraction order $j$ and the $gain(j)$ is the spectral average throughput in order $j$. NOMAD SO channel throughput has been characterised by NASA PSG team at NASA Goddard and is available on the PSG website. \cite{Liu:19} states that by considering only orders close to the central order $m$, the  $gain(j) \approx 1$ for all orders and therefore $R(A, \mathbf{\nu_{m}})$ can be rewritten as, 

\begin{equation}
  R(A, \mathbf{\nu_{m}}) = \sum_{j=m - \Delta m}^{m + \Delta m} AOTF (A,  \mathbf{\nu_{j}})\cdot F_{\textup{blaze}}(j,\mathbf{\nu_{j}})\cdot R(j,\mathbf{\nu_{j}}). 
\end{equation}

$R(j,\mathbf{\nu_{j}})$ encompasses all the \textit{signal terms}, e.g. depending on observation type, planetary surface temperature, reflectance of the surface, aerosol extinction (in particular Martian dust and water ice) and scattering. With regard to $\textup{CH}_{4}$ retrievals NOMAD SO channel diffraction order 134 is the order of most interest here, and so in this we are most interested in calculating $R(A, \mathbf{\nu_{134}})$. As a side note, it is important to distinguish $R(j,\mathbf{\nu_{j}})$ from $R(A, \mathbf{\nu_{m}})$. 

\subsection{Geometry}

In order to model NOMAD SO observations correctly, a module for computing a set of atmospheric-layer line of sight (LOS) intersection lengths is included in Ares. \cite{Tho:16} use three types of Mars shape models to calculate the aforementioned geometric parameters, these shape models include \texttt{Ellipsoidal}, \texttt{Areoid} and \texttt{Surface}. The ellipsoidal model is the most basic shape model used. Here, Mars is modelled as a tri-axial ellipsoid of radii: $3396.19 \ \textup{km} \times  3396.19\textup{km} \times 3376.2 \ \textup{km}$.  The areoidal model is not yet employed in Ares, due to the challenge of calculating the geodesic of the Areoid between (\texttt{SubObsLat},  \texttt{SubObsLon}) and (\texttt{Point0},\texttt{Point0}).

For the areoid model, Mars is modelled to a \textit{sea level}, where the gravitational and rotational potential is constant across the entire surface. This \textit{zero level} is defined by MGS/MOLA (Mars Global Surveyor)/(Mars Orbiter Laser Altimeter) at a resolution of 4 pixels per degree (around 1.6km). The surface model is the real surface elevation, calculated from a Digital Shape Kernel (DSK) by MGS/MOLA, again, at a resolution of 4 pixels per degree using \cite{Lem:01}. 

In solar occultation mode, the geometry is defined at the tangent point. The tangent point is the point on the Mars ellipsoid closest to the line of sight vector of each point. Each NOMAD shape model has a different tangent height variable, the following variables maintain NOMAD EAICD nomenclature, and are given by,  

\begin{enumerate}
    \item \texttt{TangentAlt}, is the height above the reference ellipsoid.
    \item \texttt{TangentAltAreoid}, is the height above the areoid. 
    \item \texttt{TangentAltSurface}, is the height above the surface shape model.
\end{enumerate}

Given the sublatitudes and sublongitudes of the TGO and tangent height, we are able to define an array of sublatitude and sublongitude points that define the path of solar irradiance to NOMAD. The shortest path on a spheroid, oblate spheroid, tri-axial ellipsoid and an areoidal surface will be different from one another. This said, we are not trying to find the shortest path on these surfaces, instead we are trying to find the intersection of the tangent point to spacecraft LOS with a set of spheroidal, oblate spheroidal, triaxial ellipsoidal and an areoidal surfaces.

In order to calculate the set of atmospheric-layer LOS intersection lengths, an altitude referencing system is employed. Again, maintaining EAICD nomenclature, the TGO altitude variables are given by,

\begin{enumerate}
    \item \texttt{SurfaceRadius}, is the height of the surface model above the centre of Mars.
    \item \texttt{SurfaceAltAreoid}, is the height of the surface above the reference areoid.
    \item \texttt{ObsAlt}, is the range of the spacecraft from the centre of Mars. 
\end{enumerate}

The  altitude of TGO above Mars' centre, is referenced in the NOMAD EAICD as \texttt{ObsAlt}, with the ellipsoidal, areoid and surface shape models sharing a common centre. For all geometry cases the following geometric attributes are set. 

\begin{enumerate}
    \item Assign the \texttt{Geometry} attribute for the TGO's observational altitude using \texttt{ObsAlt}.
    \item Assign the \texttt{Point0} attribute for the tangent point altitude to the Martian surface using \texttt{TangentAltSurface}.
    \item Assign the \texttt{Point0} attribute for the radius of the Martian surface relative to Mars' centre \texttt{SurfaceRadius}.
    \item Assign the \texttt{Geometry} attribute for the TGO's \\ sub-observation latitude point using \texttt{SubObsLat}.
    \item Assign the \texttt{Geometry} attribute for the TGO's \\ sub-observation longitude point using \texttt{SubObsLon}.
    \item Assign the \texttt{Point0} attribute for the sub-latitude point for the tangent point, \texttt{Lat}.
    \item Assign the \texttt{Point0} attribute for the sub-longitude point for the tangent point, \texttt{Lon}.
\end{enumerate}

$\texttt{Point0}$ corresponds to the centre of the entire field of view of a NOMAD SO bin. A longitude and latitude grid of $n_{\textup{lat-lon}}$ points is set; in this study $n_{\textup{lat-lon}}=10,000$. A fixed number of atmospheric layers, $n_{\textup{layers}}$ is also set such that $n_{\textup{layers}}=100$. Subsequently, the spacecraft's coordinates, relative to the centre of Mars, are calculated from,

\begin{align}
    x_{\textup{s}} &= r_{\textup{s}}\cos (\varphi _{\textup{s}})\sin (\theta_{\textup{s}}),\\
    y_{\textup{s}} &= r_{\textup{s}}\sin (\varphi _{\textup{s}})\sin (\theta_{\textup{s}}), \\
    z_{\textup{s}} &= r_{\textup{s}}\cos(\varphi _{\textup{s}}). 
\end{align}

$r_{\textup{s}}$ is the TGO's observation altitude, $\texttt{ObsAlt}$, relative to Mars' centre, $\varphi_{\textup{s}}$ is the TGO's $\texttt{SubObsLon}$ and $\theta_{\textup{s}}$ is the TGO's $\texttt{SubObsLat}$. Similarly the tangent point coordinates, $x_{\textup{t}}$, $y_{\textup{t}}$ and $z_{\textup{t}}$, relative to Mars' centre, are calculated as, $x_{\textup{t}} = r_{\textup{t}}\cos (\varphi_{\textup{t}})\sin (\theta_{\textup{t}})$, $y_{\textup{t}} = r_{\textup{t}}\sin (\varphi_{\textup{t}})\sin (\theta_{\textup{s}})$, and $z_{\textup{t}} = r_{\textup{t}}\cos(\varphi_{\textup{t}})$. $r_{\textup{t}}$ is the tangent point's altitude,  ($\texttt{TangentAltSurface}$ + $\texttt{SurfaceRadius}$), relative to the centre of Mars, $\varphi_{\textup{t}}$ is the tangent point's sub-longitude, $\texttt{Lon}$, and $\theta_{\textup{t}}$ is the tangent point's sub-latitude, $\texttt{Lat}$. Using Ares, a linear array of $n_{\textup{lat-lon}}$ coordinates ($x_{i}$, $y_{i}$, $z_{i}$) are calculated using the tangent point and spacecraft coordinates ($x_{\textup{t}}$, $y_{\textup{t}}$, $z_{\textup{t}}$) and ($x_{\textup{s}}$, $y_{\textup{s}}$, $z_{\textup{s}}$). A linear array of distances from the centre of Mars are calculated $\forall \ i$ as, 

\begin{equation}
    r_{i} = \sqrt{{x_{i}^{2}}+{y_{i}^{2}}+{z_{i}^{2}}}.
\end{equation}

The complexity of this challenge depends on the geometry assumed for the atmospheric layers. The following sections consider spherical, ellipsoidal and areoidal models. 

\subsubsection{Spherical}

For the spherical case, the radius of  Mars is set to $r_{\textup{Mars}}$ = 3376.20 km. Iterating over $n_{\textup{layers}}$ atmospheric layers, the goal becomes finding the points at which the LOS of NOMAD SO, defined by ($x_{\textup{t}}$, $y_{\textup{t}}$, $z_{\textup{t}}$) and ($x_{\textup{s}}$, $y_{\textup{s}}$, $z_{\textup{s}}$) intersects with the set of spherically defined atmospheric layers, ($x_{j}$, $y_{j}$, $z_{j}$), and subsequently the distances between these points, $dl_{j}$. The path lengths, $dl_{j}$ are then found by implementing an \textit{intersection criteria}. These $dl_{j}$ are subsequently passed to the Ares forward model to perform the radiative transfer integral. 

\subsubsection{Ellipsoidal}

The ellipsoid, also called a tri-axial ellipsoid is a quadratic surface, which has a general equation given by,

\begin{equation}
\frac{x^{2}}{a^{2}} + \frac{y^{2}}{b^{2}} +  \frac{z^{2}}{a^{2}}  = 1. 
\end{equation}

\noindent where for NOMAD $a=3396.19$km, $b=3396.19$km and $c=3376.20$km. For the ellipsoidal case, the Mars' polar radius is set to $r_{\textup{Mars,p}}=3376.20$ km and its equatorial radius to $r_{\textup{Mars,e}}=3396.19$ km. Analytically, we can define a set of ellipsoidal atmospheric layers, however the flattening ratio of the ellipsoid, of which the ellipsoidal atmospheric layers are a subset, should be maintained. The flattening ratio of the ellipsoid is given by, 

\begin{equation}
    f = \frac{r_{\textup{Mars,e}}-r_{\textup{Mars,p}}}{r_{\textup{Mars,e}}}.
\end{equation}

Therefore the set of ellipsoids that define the ellipsoidal atmospheric layers are given by, 

\begin{align}
    a_{i} &= 3.39619 \times 10^{6} + \Delta z_{i},\\
    b_{i} &= 3.39619 \times 10^{6} + \Delta z_{i},\\
    c_{i} &= (3.39619 \times 10^{6} + \Delta z_{i})(1-f).
\end{align}

$\Delta z_{i}$ is derived from Ares \texttt{altitudeProfile} and represents the height of atmospheric layer $i$ above the surface ellipsoid. Iterating over $n_{\textup{layers}}$ atmospheric layers, the goal becomes finding the points at which the LOS of NOMAD SO, defined by ($x_{\textup{t}}$, $y_{\textup{t}}$, $z_{\textup{t}}$) and ($x_{\textup{s}}$, $y_{\textup{s}}$, $z_{\textup{s}}$) intersects with the set of ellipsoidal defined atmospheric layers, ($x_{j}$, $y_{j}$, $z_{j}$), and subsequently the distances between these points, $dl_{j}$. The $dl_{j}$ are then found by implementing an \textit{intersection criteria} and then passing this to the Ares forward model. Ares uses the \texttt{geographiclib} package \citep{Kar:13} to calculate the ellipsoidal geodesic between ($x_{\textup{t}}$, $y_{\textup{t}}$, $z_{\textup{t}}$) and ($x_{\textup{s}}$, $y_{\textup{s}}$, $z_{\textup{s}}$), utilising \texttt{Geodesic}, \texttt{InverseLine} and \texttt{Position}.

\begin{sloppypar}
As stated previously, for the \texttt{Areoid} shape model, Mars can be modelled to a \textit{sea level}, where the gravitational and rotational potential is constant across the entire surface. The areoidal model is not yet employed in Ares, due to the challenge of calculating the geodesic of the Areoid between (\texttt{SubObsLat},  \texttt{SubObsLon}) and (\texttt{Point0},\texttt{Point0}). This requires utilising  \texttt{TangentAltAeroid}, \texttt{TangentAltAeroid}, \texttt{SurfaceRadius}  and \texttt{SurfaceAltAreoid} attributes, with MGS/MOLA data. An approximation of the geodesic required could be achieved by calculating the geodesic, as is the case for \texttt{Spherical} or  \texttt{Ellipsoidal} shape models and then mapping those latitude and longitude points to the Areoid.
\end{sloppypar}

\subsection{Chemistry}

\subsubsection{Opacities}

\begin{sloppypar}
Ares utilises the spectroscopic line lists from the HITRAN 2016  (High Resolution Transmission) database \citep{Gor:16}  to generate Martian absorption cross-sections. Through the Hitran Application Programming Interface (HAPI) \citep{Koc:19} , high resolution  absorption cross-sections with \texttt{WavenumberStep} = 0.001 $\textup{cm}^{-1}$, have been generated for the Martian atmosphere for a range of NOMAD target species. Ares absorption cross-sections include,  $\textup{C}_{2}\textup{H}_{4}$, $\textup{H}_{2}\textup{CO}$, $\textup{H}_{2}\textup{S}$,, $\textup{HO}_{2}$, $\textup{NH}_{3}$, $\textup{NO}_{2}$, OCS, $\textup{O}_{3}$, $^{12}\textup{CH}_{4}$, $^{13}\textup{CH}_{4}$, $\textup{H}_{2}\textup{O}$, $^{13}\textup{CO}_{2}$ and $^{12}\textup{CO}_{2}$. These absorption cross-sections account for expected Martian pressures, temperatures and $\textup{CO}_{2}$ broadening, due to the $\textup{CO}_{2}$-rich Martian atmosphere. Using HAPI, we generate absorption cross-sections as a function for each molecule, isotopologue, pressures and temperatures for the required wavenumber range. Ares absorption cross-sections range in temperature and pressure from 100-300 K, increasing in increments of 10 K, and 0-600 Pa, increasing in increments of 10 Pa, respectively.
\end{sloppypar}

\subsubsection{Mars Climate Database}

The Mars Climate Database (MCD) is an output dataset derived from Global Climate Model (GCM) simulations of the Martian atmosphere using Mars orbiter data assimilation \citep{Forg:99,Mill:18}. In this study, the MCD, is used to provide atmospheric priors for use in Ares forward model simulations. In this study these atmospheric priors consist of constant vertical mixing ratio profiles for $\textup{CO}_{2}$, and $\textup{H}_{2}\textup{O}$. The MCD does not yet provide atmospheric data for $\textup{CH}_{4}$, and so the vertical mixing ratio profile for $\textup{CH}_{4}$ is obtained for two cases provided by SAM-TLS. 7.2 $\pm$ 2.1 ppbv, from \cite{Web:15}, and has been adopted for the \textit{high} methane concentration case and 0.41 $\pm$ 0.16 ppbv, from \cite{Web:18}, and for the \textit{low} methane concentration case, representing the background concentration at Gale Crater.

\section{Forward Modelling}

In this section, we introduce the Ares transmission forward model, based on \cite{Wal:15} and the \texttt{Tau} model of \cite{Hol:13}. We link the Ares forward model to the aforementioned models; the NOMAD \texttt{Instrument} model, including \textit{blaze and AOTF functions}, \texttt{NomadNoise}, NOMAD \texttt{Geometry} and the Mars \texttt{Chemistry} components, namely Martian absorption cross-sections and MCD atmospheric priors. 

Forward modelling, is used with Inverse Modelling to obtain the best estimate of atmospheric properties. In the field of atmospheric sciences, these best estimates are known as atmospheric retrievals. With regard to atmospheric retrievals, the forward model solves the Radiative Transfer Equation (RTE), \cite{Cha:60}, and defines the relationship between the so called state vector, $\textbf{x}$, and measurement vector, $\textbf{y}$, given by,

\begin{equation}
\label{equ:forward}
\mathbf{y} = \mathbf{F(x,b)} + \mathbf{\epsilon}.
\end{equation}

In the aforementioned equation $\textbf{b}$ is a parameter vector that includes all the forward model parameters that we do not seek to optimize. In the case of Ares, this could include Mars' surface emissivity, or $\textup{H}_{2}\textup{O}$ and $\textup{C}\textup{O}_{2}$ ice cloud densities. $\mathbf{\epsilon}$ is the forward model error and defines the error in a simulated measured signal due to the forward model, $\textbf{F(x,b)}$. 

Inversion of $\textbf{F(x,b)}$, can then be used to obtain a statistical estimate of the state vector, $\textbf{x}$, given measurement vector $\textbf{y}$. Where $\textbf{x}$ represents the atmospheric priors and the measurement vector $\textbf{y}$, represents a set of NOMAD SO channel transmission spectra. 

If $\mathbf{\epsilon} \neq  \mathbf{0}$, then there exists an error in the Ares forward  model, the model parameters or the observations. Under all realistic scenarios $\mathbf{\epsilon} \neq  \mathbf{0}$, and so the best estimate of the state vector that can be obtained, is a statistical estimate. 

\subsection{Beer-Bouguer-Lambert Law}

The monochromatic intensity of radiation passing through a gas, $I_{\lambda }(z)$, is given by the \textit{Beer-Bouguer-Lambert Law} as a function of atmospheric altitude $z$, 
\begin{equation}
    I_{\lambda }(z) = I_{\lambda }(0) \textup{e}^{-\tau _{\lambda }(z)}.
\end{equation}

\noindent where $\lambda$ is the wavelength of the radiation, $I_{\lambda }(0)$ the radiation intensity at the top of the atmosphere and $\tau _{\lambda }(z)$ the optical depth of the medium. For a given absorbing molecular species $m$ we can define the optical depth to be the integral of the absorption cross-section $\zeta_{m}(\lambda)$, the column density $\chi_{m}(z)$, and $\rho_{N}(z)$ the number density, over the optical path length $l(z)$, given by, 

\begin{equation}
    \tau _{\lambda ,m}(z) = 2\int_{0}^{l(z)}\zeta_{m}(\lambda)\chi_{m}(z)\rho_{N}(z)dl.
\end{equation}

The optical path length $l(z)$ is defined by the geometry of the transmission of radiation through the atmosphere. In the case of Mars $l(z)$ is provided by the Ares \texttt{Geometry} module. The total optical depth is given by the sum of the individual molecular species optical depths, 

\begin{equation}
    \tau _{\lambda }(z) = \sum_{m=1}^{N_{m}}\tau _{\lambda ,m}(z),
\end{equation}

\noindent $N_{m}$ being the total number of absorbing molecular species. The monochromatic 
transmittance $\mathcal{T}_{\lambda}$ is therefore given by, 

\begin{equation}
    \mathcal{T}_{\lambda} = \textup{e}^{{\tau _{\lambda }(z)}}.
\end{equation}

Therefore calculating $\mathcal{T}_{\lambda}$ for all $\lambda$ points in the spectral region of interest, e.g. the bandwidth of NOMAD SO channel diffraction order 134, with a \textit{transmittance equivalent} of $R(A, \mathbf{\nu_{m}})$ enables the simulation of NOMAD SO channel transmission spectra, 

\begin{equation}
  \mathcal{T}(A, \mathbf{\nu_{m}}) = \frac{\sum_{j=m - \Delta m}^{m + \Delta m} AOTF (A,  \mathbf{\nu_{j}}) F_{\textup{blaze}}(j,\mathbf{\nu_{j}}) gain(j) \mathcal{T}(j,\mathbf{\nu_{j}})}{PEC(A)}. 
\end{equation}

For further details on forward modelling we refer the reader to \cite{Wal:15} and \cite{Hol:13}. 

\section{Analysis of simulated NOMAD observations with Ares}

\subsection{Bayesian analysis of planetary atmospheres}

The Ares framework allows the user to select a range of optimisation routines to fit the atmospheric forward model to the data. In planetary science retrievals, the use of optimal estimation \citep{Rod:00} is commonplace. Optimal Estimation (OE) calculates the maximum likelihood (ML) through a regularised matrix inversion of a Gaussian likelihood given prior constraints. OE methods yield excellent results in high signal-to-noise data regimes with good \textit{a priori} knowledge of expected atmospheric forward-model parameters. However, in the case of low signal measurements and poorly constrained prior ranges, ML methods  often insufficiently capture possible correlations in the likelihood surface for under-constrained retrievals. A typical example is atmospheric retrieval of extra-solar planet atmospheres where low-resolution data leads to significant correlations between measured trace-gas abundances. In these cases, it is necessary to solve for the full posterior probability distribution of the Bayesian argument. 

Similarly, measuring trace-gas abundances at the detection limits of current instrumentation in the Martian atmosphere will require a complete understanding of possible correlations in the posterior distributions of the retrieval solution. 

Following the definition of equation\,\ref{equ:forward}, we can state Bayes' Theorem as, 

\begin{equation}
P(\mathbf{y}| \mathbf{x},\mathcal{M}) = \frac{P(\mathbf{x}| \mathbf{y},\mathcal{M})P(\mathbf{y},\mathcal{M})}{P(\mathbf{x},\mathcal{M})}.
\label{equ:bayes}
\end{equation}

$\mathbf{x}$ is the observed spectrum (e.g. the NOMAD SO channel measurement), $\mathbf{y}$ is the vector of atmospheric forward model parameters, and $\mathcal{M}$ is the atmospheric forward model used. Hence, we want to compute the posterior distribution $P(\mathbf{y}| \mathbf{x},\mathcal{M})$, describing the probability distribution of forward model parameters for a given spectrum and forward model. By default, we assume the likelihood, $P(\mathbf{x}| \mathbf{y},\mathcal{M})$ to follow a Normal distribution, 

\begin{equation}
P({\bf x} | {\bf y}, \mathcal{M}) = \frac{1}{{\bf \varepsilon} \sqrt{2\pi}} ~\text{exp}\left [{-\frac{1}{2}\sum_{\nu}^{N}\left ( \frac{x_{\nu} - {\bf F}_{\lambda} ({\bf y, b}) }{\varepsilon_{\nu}} \right )^{2}} \right ],
\label{equ:normal-likelihood}
\end{equation}

\noindent though other distirbutions can be specified in Ares. 
The prior distribution $P(\mathbf{y},\mathcal{M})$ encapsulates the prior information for a given parameter and atmospheric model. In the case of Optimal Estimation, this prior must be Gaussian. In the Ares framework, the prior distribution can be any continuous probability distribution, allowing for informative as well as uninformative priors (e.g. Jeffrey's priors). By default, we utilise uninformative log-uniform prior distributions over retrieved trace-gas abundances to ensure a maximally data driven retrieval solution. Finally, we define the Bayesian Evidence as the integral of the likelihood and prior:

\begin{equation}
\label{equ:bayesevidence}
P(\mathbf{x},\mathcal{M}) = \int P({\bf y}, \mathcal{M}) P({\bf x} | {\bf y}, \mathcal{M}) \text{d}{\bf y}.
\end{equation}

The evidence allows for Bayesian model selection, i.e. the statistical comparison of the atmospheric with that of another. A commonly used scale for model significance is given by \cite{Kas:95}.

\subsection{Maximum a posteriori (MAP) Solution} 

The $\textit{maximum a posteriori}$ (MAP) solution is given by the maximum possible value of $P(\mathbf{y}| \mathbf{x},\mathcal{M})$. The maximum of \\ $P(\mathbf{y}| \mathbf{x},\mathcal{M})$ is given by, 

\begin{equation}
\nabla_{\mathbf{y}} P(\textbf{y}|\textbf{x},\mathcal{M}) = \frac{\partial }{\partial \textbf{y}}P(\textbf{y}|\textbf{x},\mathcal{M}) =\textbf{0}.
\end{equation}

$\nabla_{\mathbf{y}} = \frac{\partial }{\partial \textbf{y}}$ is the gradient operator in the state vector space and $\textbf{0}$ is the zero vector in the state vector space. For uniform prior distributions, the MAP is equivalent to the maximum likelihood, and hence the result of Optimal Estimation. Whilst the MAP solution is informative, it does not reflect the full inter-parameter correlation possible in under constrained likelihoods. Similarly, by their definition, MAPs do not carry possibly diagnostic information on local maxima in the posterior distribution. By mapping the full posterior, we are able to derive a more robust understanding of the retrievability of trace gases at very low volume mixing ratios.  

\subsection{MCMC} 

Monte Carlo Markov Chain (MCMC) methods are often implemented in the field of exoplanetary science, whereby \\ MCMC routines explore the likelihood space by means of a Markovian chain. TauREx3 provides an implementation of delayed rejection (DR) Adaptive MCMC (DRAM, \cite{Haa:06}). For further details we refer the reader to the literature; for brevity we provide a brief overview here. DRAM differs from the classical Metropolis-Hastings sampler \cite{Met:53}, \cite{Has:70}, \cite{Brooks:11}, in that a DR algorithm is implemented and that an adaptive proposal distribution is calibrated using the co-variance of the sample path of the MCMC chain. MCMC does not typically sample the full likelihood volume and hence does not allow us to calculate the Bayesian evidence. Hence equation\,\ref{equ:bayes} is approximated as 
\begin{equation}
    P(\mathbf{y}| \mathbf{x}, \mathcal{M}) \propto  P(\mathbf{x}| \mathbf{y},\mathcal{M})P(\mathbf{y},\mathcal{M}).
\end{equation}

For implementation we refer the reader to Appendix A of Waldmann et al. 2015. TauREx3 runs several parallelised MCMC chains, in order to check convergence and increase the sampling of the likelihood space of possible solutions.

\subsection{Nested Sampling} 

Nested Sampling (NS) algorithms \cite{Ski:06} are frequently used in exoplanetary science for performing atmospheric retrievals and constitute the standard sampling method in Ares, \texttt{Nestle}, \cite{Bar:15} and \texttt{MultiNest}, \cite{Fer:09}. Whilst MCMC methods explore the entire likelihood space by means of a Markov chain, NS applies a general Monte Carlo analysis to periodically constrain ellipsoids that encompass spaces of highest likelihood. Through NS, the Bayesian partition function (i.e. the Bayesian evidence, \textit{E}) can be calculated, allowing for exact model selection and a complete sampling of the likelihood. NS is the default sampler for the Ares framework. 

\section{Ares - Planetary Spectrum Generator Comparison} 

\begin{figure*}[!hbtp]
\centering
  \includegraphics[width=16cm]{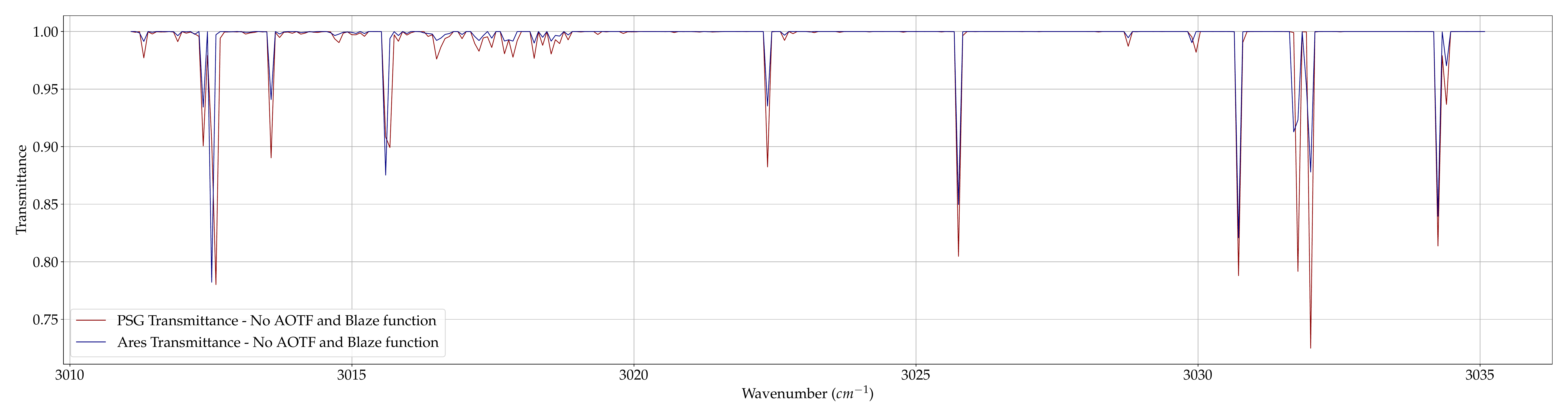}\\
  \caption{Ares PSG NOMAD SO channel solar occultation transmittance simulation comparison, covering the spectral range of diffraction order 134, 3011-3035 $\textup{cm}^{-1}$, using mutual input variables, such as geometry and atmospheric priors. NOMAD AOTF and blaze function are not applied in this figure. Atmospheric priors and prior limits are provided in the top section of Table 1.}\label{fig7}
\end{figure*}

\begin{figure*}[!hbtp]
\centering
  \includegraphics[width=16cm]{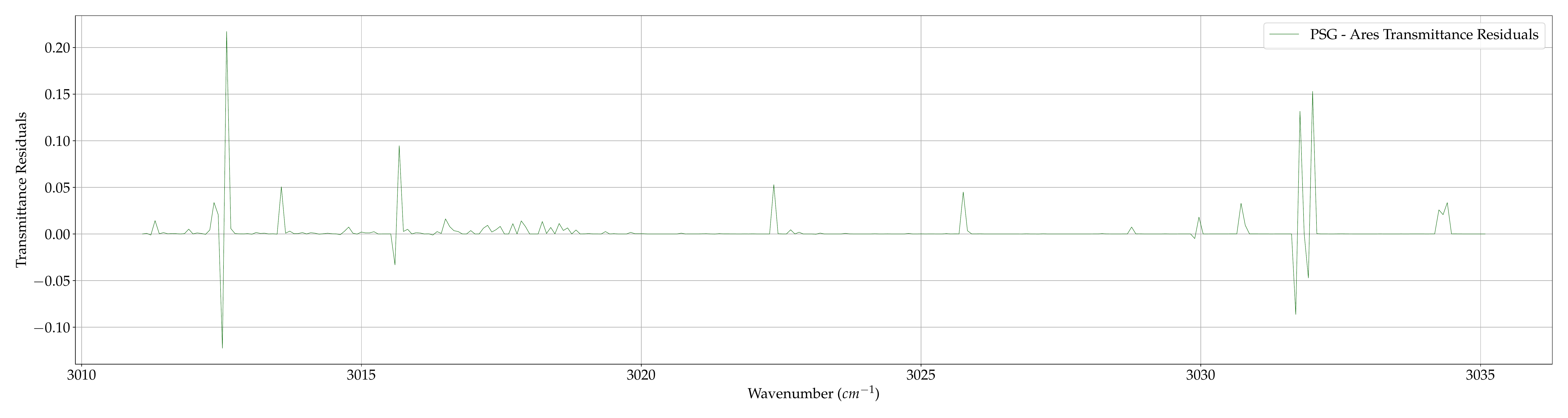}\\
  \caption{Ares PSG NOMAD SO channel solar occultation transmittance simulation residuals for Figure \ref{fig7}.}\label{fig8}
\end{figure*}

\subsection{Validation} 

In order to validate Ares, the Ares forward model has been compared against NASA Goddard's Planetary Spectrum Generator (PSG), \cite{Vil:18}. The PSG is an extensive online radiative transfer suite that is capable of synthesising Martian NOMAD SO channel spectra. Like Ares, the PSG can utilise atmospheric priors from the MCD, \cite{Forg:99}, \cite{Mill:18}. Ares Martian transmittance and spectral radiance spectra, have been simulated using NOMAD observation metadata and compared against the PSG; running Ares and PSG forward models with mutual input variables. 

This comparison is achieved through firstly populating PSG API \textit{config} files with the same NOMAD observation metadata and priors, as is used for the Ares forward model simulations, such as tangent height, TGO observation altitude and atmospheric mixing ratio profiles. Following the population of the PSG configuration files a set of \texttt{curl} commands using the aforementioned \textit{config} files are sent to the PSG API, for example:

\begin{verbatim}
curl -d type=rad -d whdr=n --data-urlencode
file@config.txt https://psg.gsfc.nasa.gov/api.php 
-o config_output.txt
\end{verbatim}

Setting $\texttt{type=trn}$ in the curl command, with the radiation unit in the $\textit{config}$ file set to $\texttt{rif}$, returns the PSG transmittance simulation. Similarly, setting $\texttt{type=rad}$ with the radiation unit in the $\textit{config}$ file set to $\texttt{Wsrm2cm}$ returns the PSG spectral radiance simulation. Ares, unlike the PSG, does not yet correct for Instrument Line Shape (ILS) ghosts, this is left to future work. The \textit{config} files and \textit{config-generator} associated with this comparison will be available in the Supplementary Material.

In all simulations, the vertical mixing ratio and temperature profiles are assumed to be constant and isothermal respectively. In Ares, the Sun is assumed to be a blackbody, with $T_{\odot }=5778$ K. The wavenumber range of all simulations covers $\nu_{\textup{min}}=2945 \ \textup{cm}^{-1}$ to $\nu_{\textup{min}}=3130 \ \textup{cm}^{-1}$, focusing on  $\nu_{\textup{min}}=3011 \ \textup{cm}^{-1}$ to $\nu_{\textup{min}}=3035 \ \textup{cm}^{-1}$, in order to cover NOMAD SO channel diffraction order 134, containing the $\textup{CH}_{4}$ $\nu_{3}$ band. The minimum and maximum atmospheric pressures are set to $p_{\textup{min}}$ = 1 Pa and $p_{\textup{max}}$ = 600 Pa respectively. In this initial comparison we do not include scattering or collision induced absorption. 

\subsection{AOTF and Blaze function comparison} 

Whilst a transmittance simulation comparison between the PSG and Ares is possible, see Figure \ref{fig7} and \ref{fig8}, a direct transmittance comparison with the NOMAD SO channel AOTF and blaze function applied is not. This is due to the PSG AOTF and blaze function transformations being inapplicable to transmittance simulations. Therefore, in order to correctly compare and validate Ares against the PSG, with and without the AOTF and blaze function applied, a conversion of Ares simulated transmittance to spectral radiance is required, or alternatively a conversion of the simulated PSG spectral radiance simulations, Figure \ref{fig14}, back to a transmittance simulation, Figure \ref{fig11}, after the PSG AOTF and blaze function transformations have been applied. In order to enable the AOTF and blaze function in the PSG, the generator telescope is set to $\texttt{AOTF}$. To disable the AOTF and blaze function the generator telescope is set to $\texttt{Single Dish}$, \cite{Vil:18}. 

\subsection{Geometry comparison} 

For Ares, we observed that the simulated Ares transmittance spectra with the \texttt{Ellipsoidal} and \texttt{Spherical} geometry modules matches the PSG well with the same input variables. We observe, that the residuals between the \texttt{Ellipsoidal} and the PSG are smaller than when utilising the \texttt{Spherical} module. There are minor differences in the transmittance simulation absorption depths between, Ares and PSG, which could be due to using different absorption cross-sections or due to the differences in the Ares and PSG geometry module implementations.

\begin{figure*}[!htbp]
\centering
  \includegraphics[width=16.0cm]{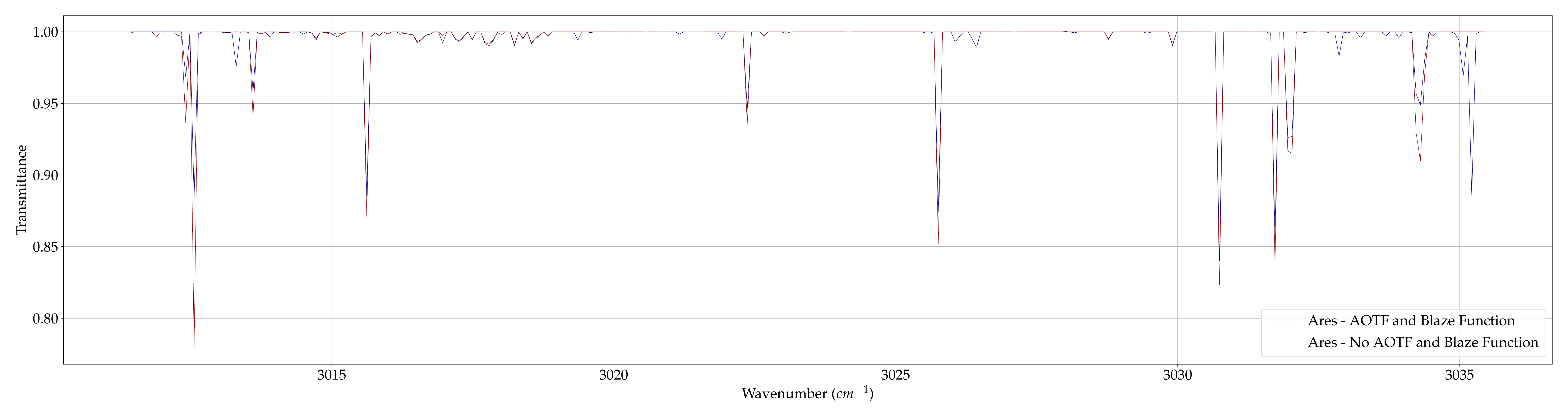}\\
  \caption{Ares NOMAD SO channel solar occultation transmittance simulation comparison, with and without AOTF and blaze function applied, covering the spectral range of diffraction order 134, 3011-3035 $\textup{cm}^{-1}$.}\label{fig9}
\end{figure*} 

\begin{figure*}[!htbp]
\centering
  \includegraphics[width=16.0cm]{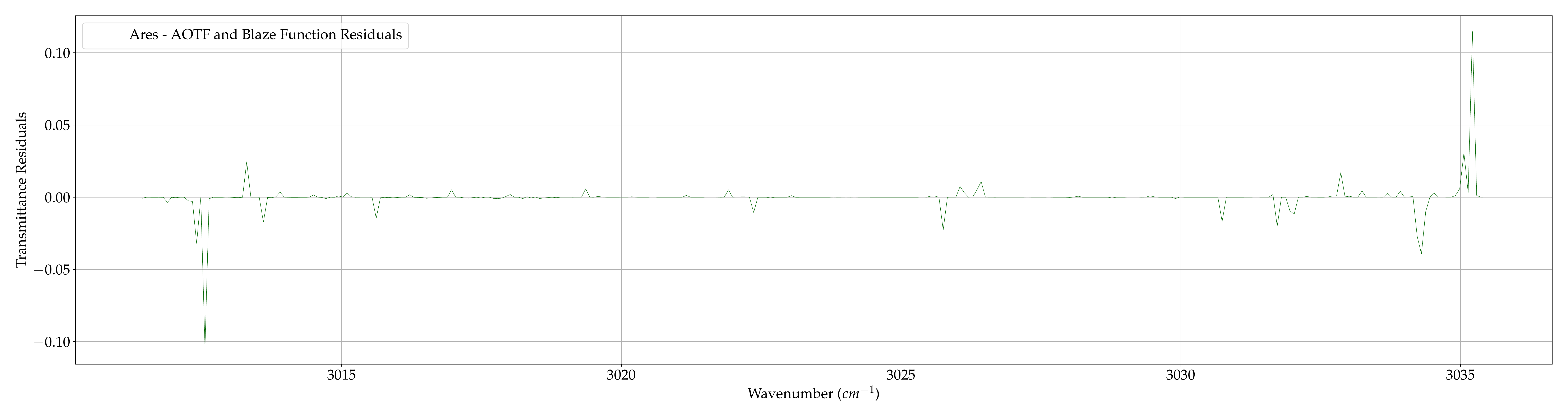}\\
  \caption{Ares NOMAD SO channel solar occultation transmittance simulation residuals for Figure \ref{fig9}.}\label{fig10}
\end{figure*} 

\begin{figure*}[!htbp]
\centering
  \includegraphics[width=16.0cm]{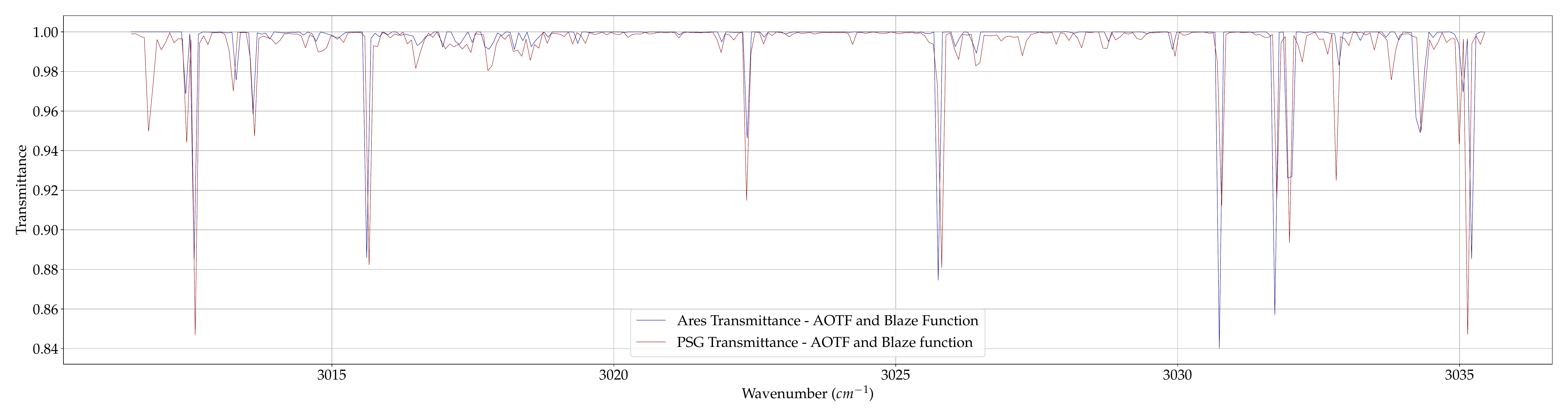}\\
  \caption{Ares PSG NOMAD SO channel solar occultation transmittance simulation comparison, with NOMAD SO channel AOTF and blaze function applied, covering the spectral range of diffraction order 134, 3011-3035 $\textup{cm}^{-1}$.}\label{fig11}
\end{figure*} 

\begin{figure*}[!htbp]
\centering
  \includegraphics[width=16.0cm]{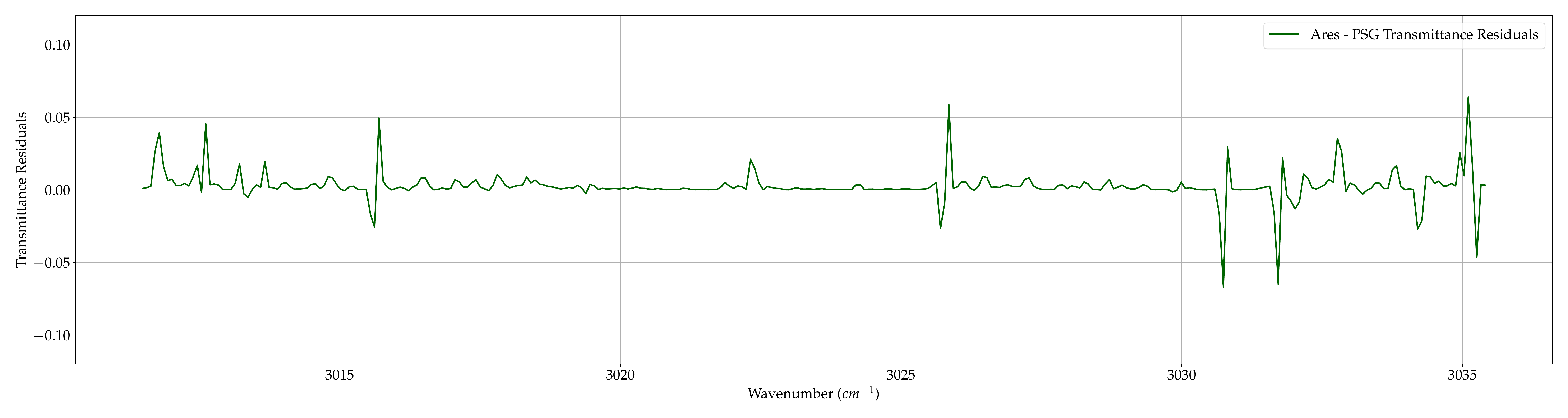}\\
  \caption{Ares PSG NOMAD SO channel solar occultation transmittance simulation residuals for Figure \ref{fig11}.}\label{fig12}
\end{figure*} 

\begin{figure*}[!htbp] 
\centering
  \includegraphics[width=17.0cm]{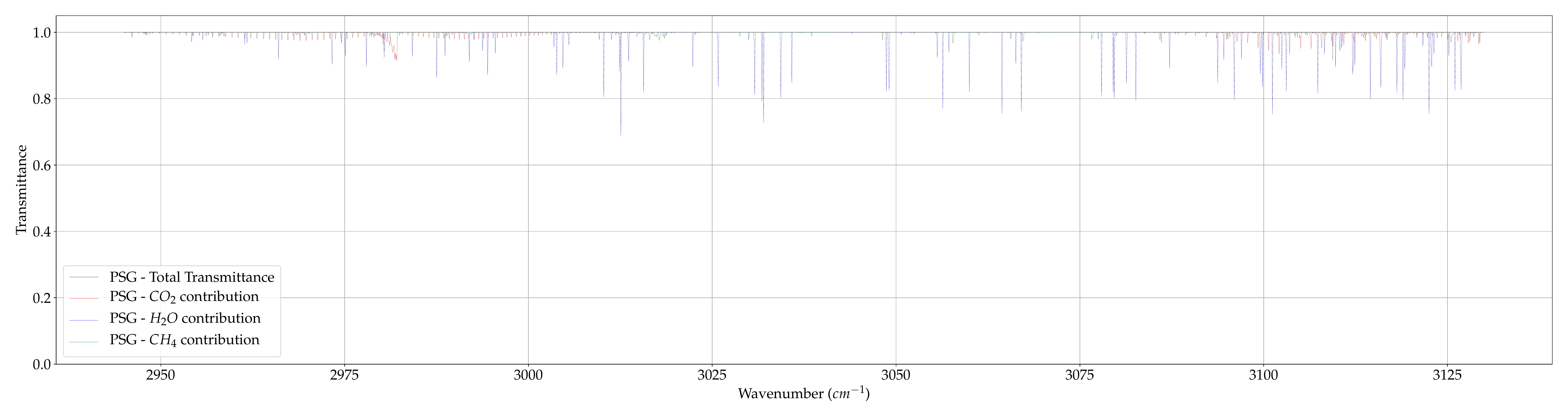}\\
  \caption{PSG NOMAD SO channel solar occultation simulated transmittance measurement for 2945-3130 $\textup{cm}^{-1}$, showing molecular contributions of $\textup{CO}_{2}$, $\textup{H}_{2}\textup{O}$ and $\textup{CH}_{4}$.}\label{fig13}
\end{figure*}

\begin{figure*}[!htbp] 
\centering
  \includegraphics[width=17.0cm]{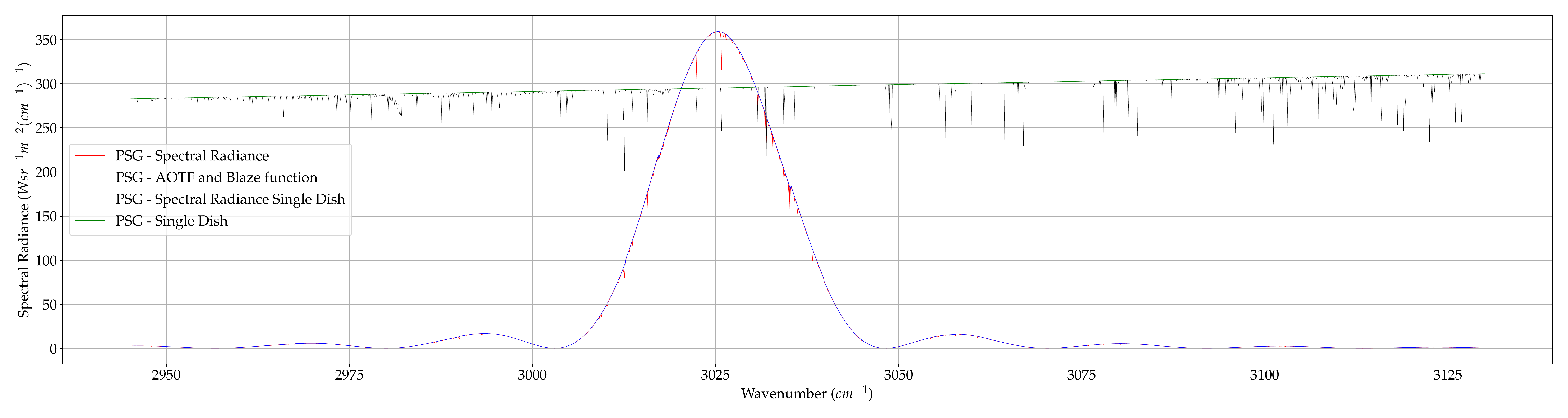}\\
  \caption{A graph of a PSG simulated spectral radiance measurement for NOMAD SO channel solar occultation for 2945 - 3130 $\textup{cm}^{-1}$, with and without NOMAD SO channel AOTF and blaze function applied.}\label{fig14}
\end{figure*}

\begin{table}[htbp]
  \caption{Ares - PSG forward model parameters and priors.}
   \label{extremos45}
  \begin{tabular*}{\hsize}{lrrrr}
\hline
    Parameter & Prior & Unit & Prior limits \\ \hline
    $\textup{H}_{2}\textup{O}$ & $10^{5}$ & ppbv & [$1.0 \times 10^{-3}$, $1.0 \times 10^{7}$] \\
    $\textup{CO}_{2}$ & $9.65 \times 10^{8}$ & ppbv & [$9.0 \times 10^{8}$, $9.8 \times 10^{8}$] \\
    T & 230 & K & [130.0,330.0] \\ 
    $\textup{CH}_{4,high}$ & 7.2 & ppbv & [$1.0 \times 10^{-3}, 1.0 \times 10^{2}$] \\ \hline
    $^{13}\textup{CH}_{4}$ & 0.072 & ppbv & [$1.0 \times 10^{-3}, 1.0 \times 10^{7}$] \\
    $\textup{OCS}$ & 1.0 & ppbv & [$1.0 \times 10^{-3}, 1.0 \times 10^{7}$] \\
    $\textup{PH}_{3}$ & 1.0 & ppbv & [$1.0 \times 10^{-3}, 1.0 \times 10^{7}$] \\
    $\textup{NH}_{3}$ & 1.0 & ppbv & [$1.0 \times 10^{-3}, 1.0 \times 10^{7}$] \\ \hline
  \end{tabular*}
\end{table}

\begin{figure*}[!hbtp]
\centering
  \includegraphics[width=15cm]{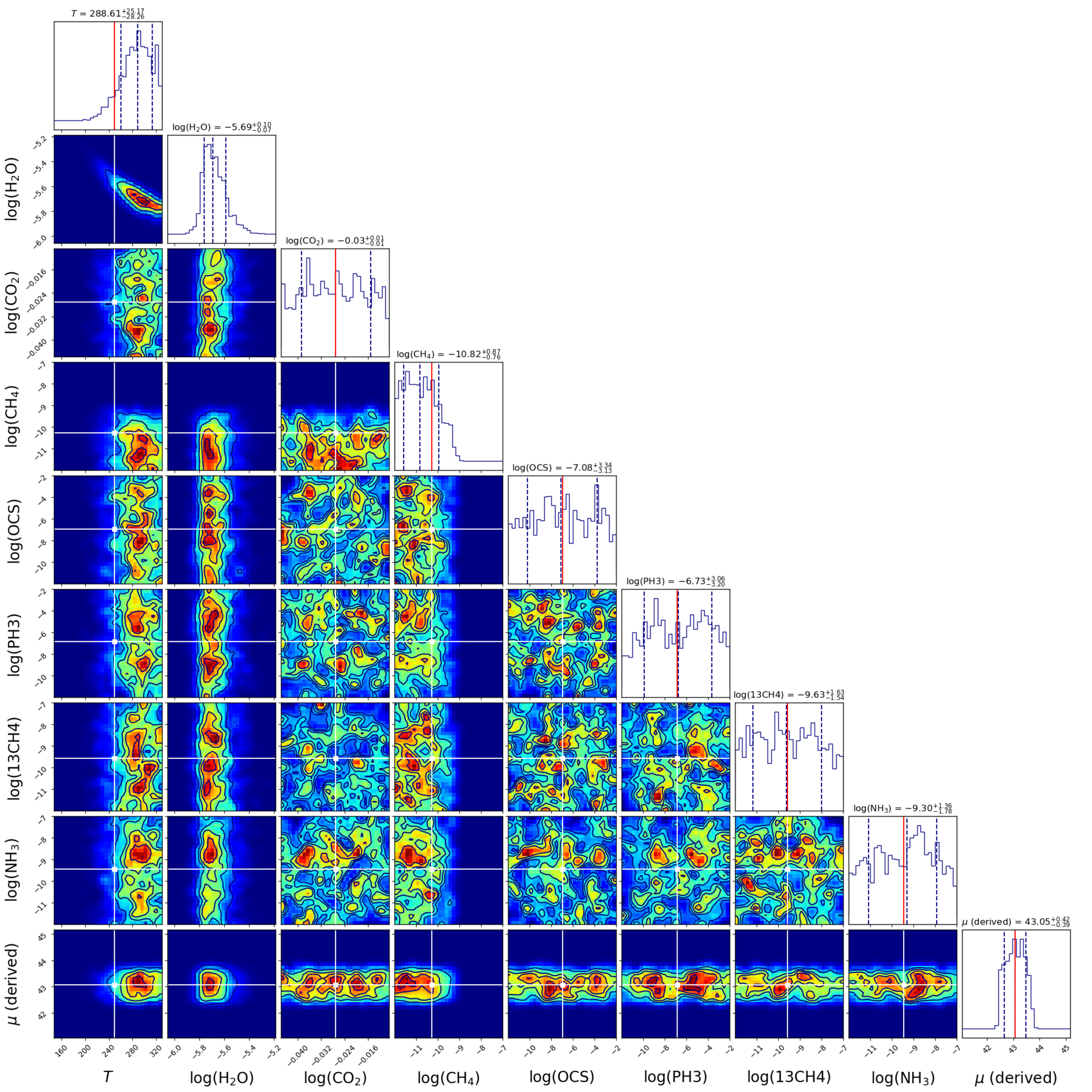}\\
  \caption{Posterior distributions of \textit{simulated} NOMAD spectrum retrieved using Ares, for diffraction order 134, using observation: 20180430-154925-1p0a-SO-A-I-134 metadata, index (78,1), tangent height 20.504 km. Contours represent regions of high (red) and low (blue) probability density. Ares priors are represented by the red line in the marginalised distributions and as white cross hairs in the joint distributions. Dashed lines in the marginalised distributions represent are within $\sigma_{1}$ of the MAP estimate. We observe that the retrievals for OCS, $\textup{PH}_{3}$ and $^{13}\textup{CH}_{4}$ are poorly constrained. As expected, we note that retrievals of $\textup{H}_{2}\textup{O}$ and temperature are correlated. Figure \ref{fig15} was produced using the \texttt{corner.py} package, \cite{FM:16}.}\label{fig15}
\end{figure*}

\begin{figure*}[!hbtp]
\centering
  \includegraphics[width=17cm]{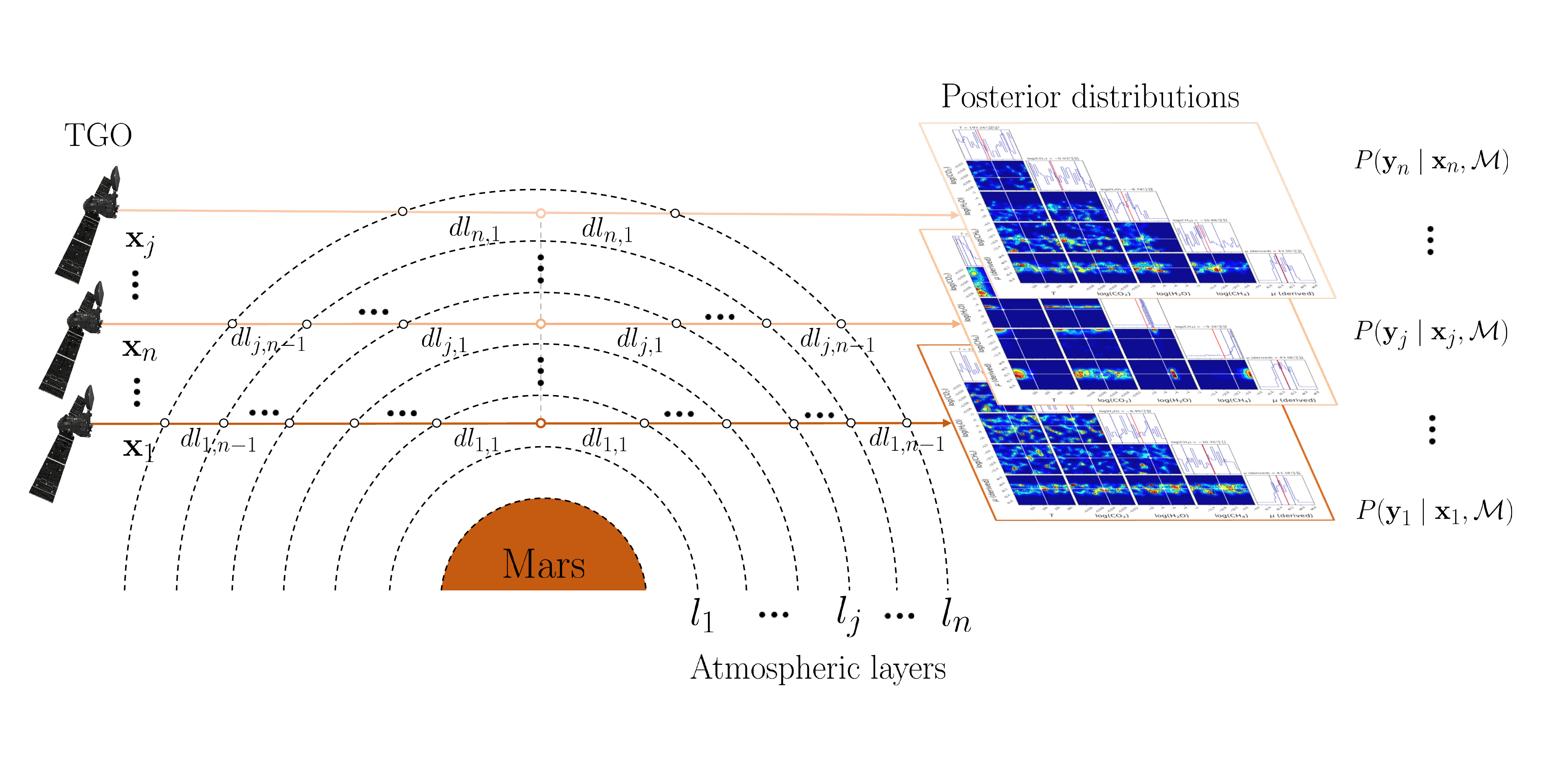}\\
  \caption{A schematic diagram of the TGO, Martian atmospheric layers and corresponding Ares a posterior distributions, mapping the exact statistical correlations between atmospheric parameters. Shown is a layer-by-layer retrieved atmosphere using simulated NOMAD spectra, for diffraction order 134, at three tangent heights, in the lower, middle and upper Martian atmosphere. We observe that the retrievals for lower tangent heights are strongly noise dominated while middle atmosphere retrievals produce tighter constraints.}\label{fig16}
\end{figure*}

\section{Results and Observations}

Validating Ares through comparison to other established forward models is critical prior to progressing to perform Ares retrievals. The results of the Ares PSG forward model comparisons show that both models are generally in good agreement. 

\subsection{AOTF and Blaze function comparison}

Figure \ref{fig7}, shows the Ares PSG NOMAD SO channel solar occultation transmittance simulation comparison, without applying AOTF and blaze function. We generally note minor differences in the absorption line depths. Figure \ref{fig8}, shows the transmittance residuals of Figure \ref{fig7}, with a maximum residual of 0.217 at 3012.59 $\textup{cm}^{-1}$ and a minimum residual of -0.123 at 3012.51 $\textup{cm}^{-1}$. Whilst these residuals are significant, the close proximity of the maximum and minimum residual, coupled with the similarity of absorption depths in Figure \ref{fig7}, suggests that these differences are due to minor discrepancies in the spectral line positions. 

Figure \ref{fig9}, shows Ares NOMAD SO channel solar occultation simulated transmittance, with and without the AOTF and blaze function applied. Figure \ref{fig10}, shows the transmittance residuals of Figure \ref{fig9}, with a maximum residual of 0.115 at 3035.21 $\textup{cm}^{-1}$ and a minimum residual of -0.105 at 3012.56 $\textup{cm}^{-1}$. Figure \ref{fig11}, shows the Ares PSG NOMAD SO channel solar occultation transmittance simulation comparison with the AOTF and blaze function applied. Figure \ref{fig12}, shows the transmittance residuals of Figure \ref{fig11}, with a maximum residual of 0.063 at 3035.11 $\textup{cm}^{-1}$ and a minimum residual of -0.067 at 3030.74 $\textup{cm}^{-1}$. We observe that applying the AOTF and blaze function, the PSG and Ares, introduces absorption features, that are not present prior to AOTF and blaze function transformations. This exercise establishes that Ares and the PSG produce comparable results when the NOMAD SO channel AOTF and blaze function are applied.

\subsection{Geometry comparison}

We note a slight difference in absorption line depth between using Ares $\texttt{Spherical}$ and $\texttt{Ellipsoidal}$ geometry modules. However, this slight variation is dependent on the tangent height. The residuals between Ares geometry modules, across the simulated waveband, appears to decrease with increasing altitude. This is to be expected, as the lower the altitude of the tangent point (or atmospheric LOS intersection point), the greater the difference between the calculated ellipsoidal and spherical path lengths. This difference, in turn is exacerbated by the increase in atmospheric density at lower altitudes. The residuals between Ares and PSG transmission spectra, across the simulated waveband, decrease with increasing altitude. This confirms the expectation, that the most challenging observations are those at lower altitudes \cite{Liu:19}. 

\subsection{Retrievals}

Following the Ares PSG comparison, Ares marginalised and conditional posterior distributions have been generated for the aforementioned Ares simulated spectra Figure \ref{fig11}, shown in Figures \ref{fig15} and \ref{fig16}. We have assigned uncertainties for each wavelength bin to give a signal-to-noise ratio of the maximum value expected by \cite{Liu:19} of 1000 and utilised Nested Sampling, with \texttt{Nestle} set as the optimizer, with 200 live points.

Figure \ref{fig15}, shows an example of a set of Ares posterior distributions, retrieved using Ares, from a simulated NOMAD SO channel transmittance measurement. This retrieval covers diffraction order 134, using TGO observation 20180430-154925-1p0a-SO-A-I-134 metadata, with index (78,1), with tangent height 20.504 km. The \texttt{jet} colour scheme of the joint distributions is used to represent differences in probability density, whereby regions of high and low probability density are represented within red and blue regions respectively. Ares priors are represented by the red line in the marginalised distributions and as white cross hairs in the joint distributions. As expected, we note that retrievals of $\textup{H}_{2}\textup{O}$ and temperature are correlated. For this particular observation we find an upper limit of methane at log$_{10}(\chi) = -10.82 _{-0.76}^{+0.87}$ and no constraint of OCS, PH$_3$, NH$_3$ and $^{13}$CH$_4$ beyond their prior bounds. We note that the tight prior bounds of CO$_2$ effectively constrain its abundance given the data and no further constraint within these priors is  retrieved. The retrieved parameters are shown in Table 2.

Figure \ref{fig16}, visualises the advantages of utilising Ares to perform retrievals of forward model parameters, showing the TGO, Martian atmospheric layers and corresponding Ares a posterior distributions, for a set of NOMAD SO channel solar occultation measurements, ${\bf x_j}$. The set of probability density functions produced, $P({\bf y}|{\bf x},\mathcal{M})$, map the statistical correlations between atmospheric forward model parameters as function of altitude. Figure \ref{fig16} shows a layer-by-layer retrieved atmosphere using simulated NOMAD spectra, for diffraction order 134, at three tangent heights, in the lower, middle and upper Martian atmosphere. We observe that retrievals for middle atmosphere tangent heights produce tighter constraints of probability density. This is expected as the optical depth is too high close to the surface and too low in the upper atmosphere. Whilst Ares is a simplified model of complex atmospheric phenomena, covering only a small subset of possible atmospheric conditions and TGO viewing geometries, it is nonetheless capable of gaining unique insight into the Martian atmosphere.

\begin{table*}[ht]
  \caption{Ares forward model priors and retrieved parameters for Figure \ref{fig16}.}
\centering
\begin{tabular}{p{0.15\linewidth}p{0.15\linewidth}p{0.15\linewidth}p{0.2\linewidth}p{0.2\linewidth}}
\hline Parameter & Prior & Unit & Prior limits & Retrieved log(Parameters) \\ \hline
    $\textup{H}_{2}\textup{O}$ & $10^{5}$ & ppbv & [$1.0 \times 10^{-3}, 1.0 \times 10^{7}$] & $-5.69 _{-0.10}^{+0.07}$ \\
    $\textup{CO}_{2}$ & $9.65 \times 10^{8}$ & ppbv & [$9.0 \times 10^{8}$, $9.8 \times 10^{8}$] & $-0.03 _{-0.01}^{+0.01}$\\
    T & 230 & K & [130.0, 330.0] & $288.61 _{-25.17}^{+28.26}$ \\ 
    $\textup{CH}_{4,high}$ & 7.2 & ppbv & [$1.0 \times 10^{-3}, 1.0 \times 10^{2}$] & $-10.82 _{-0.76}^{+0.87}$ \\ 
    $^{13}\textup{CH}_{4}$ & 0.072 & ppbv & [$1.0 \times 10^{-3}, 1.0 \times 10^{7}$] & $-9.63 _{-1.54}^{+1.63}$ \\
    $\textup{OCS}$ & 1.0 & ppbv & [$1.0 \times 10^{-3}, 1.0 \times 10^{7}$] & $-7.08 _{-3.13}^{+3.34}$ \\
    $\textup{NH}_{3}$ & 1.0 & ppbv & [$1.0 \times 10^{-3}, 1.0 \times 10^{7}$] & $-9.30 _{-1.78}^{+1.36}$ \\ 
    $\textup{PH}_{3}$ & 1.0 & ppbv & [$1.0 \times 10^{-3}, 1.0 \times 10^{7}$] & $-6.73 _{-3.20}^{+3.06}$ \\ \hline
\end{tabular}
\end{table*}  

\section{Conclusions}

This research has described a novel retrieval framework, Ares, the Mars branch of TauREx\,3 \citep{Al-R:19}, designed for TGO NOMAD SO channel solar occultation measurements, the first in a series of papers. This research has characterised the instrument capabilities of the NOMAD instrument and the TauREx\,3 based Ares model. In this research, Ares has been used to simulate transmittance and spectral radiance measurements and perform retrievals of typical Martian mixing ratio vertical profiles of $\textup{CH}_{4}$, $\textup{H}_{2}\textup{O}$, and $\textup{CO}_{2}$. The Ares forward model can simulate NOMAD spectra, including instrumental effects due to the blaze function, AOTF, as well as AOTF temperature dependence.

Ares and PSG simulated transmittance and spectral radiance measurements have been compared and are generally in good agreement, with marginalised and conditional posterior distributions of simulated and real data presented. We hope that Ares will help accelerate the use of advanced statistical sampling techniques, such as MCMC and Nested Sampling to fully explore the atmospheric forward model degeneracies encountered in low signal regimes where trace-gas abundances approach the instrument sensitivity floor. The work presented here will provide a new open-source tool to the planetary science community to study very faint spectral signatures in planetary atmospheres using a statistically robust framework.

Future forward model comparisons, made against the PSG,  NEMESIS, \cite{Ir:08}, and ASIMUT-ALVL, \cite{Van:06}, will be used to further validate and refine Ares. Beyond Mars, Ares finds applications in the spectral analysis of Titan, \cite{Bel:09}, the middle atmosphere of Venus, \cite{Lim:18}, and the plumes of Enceladus, \cite{Wai:06}. 

\section{Acknowledgements}

We would like to thank the UK Space Agency for their support of this studentship through the Aurora Science Programme, STFC number 535385. We would like to thank The Royal Belgian Institute for Space Aeronomy (BIRA-IASB) for providing test NOMAD spectra, in particular Ian Thomas for help in understanding the intricacies of NOMAD. The NOMAD experiment is led by BIRA-IASB and assisted by Co-PI teams from Spain (IAA-CSIC), Italy (INAF-IAPS), and the United Kingdom (Open University). We would like to thank the LMD for providing the MCDv5.3 and Ehouarn Millour for installation help. This project has received funding from the European Research Council (ERC) under the European Union's Horizon 2020 Research and Innovation Programme (Grant agreement No. 758892, ExoAI). Furthermore, we acknowledge funding by the Science and Technology Funding Council (STFC) grants: ST/K502406/1, ST/P000282/1, ST/P002153/1 and ST/S002634/1.

\section{Declaration of Interest Statement}
The authors declare no competing financial interests.

\label{}

\section{Supplementary Material}

Following publication of this work supplementary material associated with Ares and the Ares code will be provided open source, licensed under a BSD license, available at: \\ \texttt{https://github.com/ucl-exoplanets/ares}.

\appendix

\bibliographystyle{elsarticle-harv}

\section{NOMAD Noise Model}

Real NOMAD SO channel data will include noise, therefore in order to test the ability of Ares to perform retrievals on simulated NOMAD SO observations, a module for computing the expected noise is included. A fundamental component of Ares is the \texttt{NomadNoise} class. This class considers all the sources of noise present within NOMAD observations. As stated in \cite{Nef:15}, the noise of the channels is the sum of the different noise contributions, given by the following,

\begin{equation}
N = N_{\textup{dark}} + N_{\textup{ro}} + N_{\textup{quant}} +  N_{\textup{shot}} +  N_{\textup{tb}}.
\end{equation}

The following list explains in detail the aforementioned \\ sources of noise:

\begin{enumerate}
    \item The \textit{dark current noise}, $N_{\textup{dark}}$, is a fixed source of noise, measured in electrons per pixel per second. The assumed value of the dark current for NOMAD IR channels is 6000 $\textup{e}^{-} \textup{pixels}^{-1}\textup{s}^{-1}$. $N_{\textup{dark}}$ is given by, 
    \begin{equation}
        N_{\textup{dark}} = \sqrt{6000\Delta t}.
    \end{equation}
    where $\Delta t$ is the integration time. Integration time metadata is given with NOMAD observations. 
    
    \item The \textit{readout noise}, $N_{\textup{ro}}$, is a fixed source of noise due to the detector readout, measured in pixels per second per observation. For the MARS-MW detector, in SO and LNO channels, $N_{\textup{ro}}$ is given by,
    \begin{equation}
        N_{\textup{ro}} \approx 1000 \ \textup{e}^{-} \textup{pixel}^{-1}.
    \end{equation}
    
    \item The \textit{thermal background noise}, $N_{\textup{tb}}$, is the noise in the detector produced by the thermal background. $N_{\textup{tb}}$ is given by,
    \begin{equation}
        N_{\textup{tb}}(i)= \sqrt{S_{\textup{tb}}(i)}.
    \end{equation}
    where $S_{\textup{tb}}(i)$ represents the electrons generated in detector pixel $i$ due to thermal emission coming from the instrument. Whilst $N_{\textup{tb}}$ is challenging to simulate, \cite{Liu:19}, provides a set of expected thermal background noise values given a set of NOMAD instrument temperatures. At 263K $N_{\textup{tb}} = 1.5 \times 10^{7} \  \textup{e}^{-}\textup{s}^{-1}$, at 273K $N_{\textup{tb}} = 3.0 \times 10^{7} \  \textup{e}^{-}\textup{s}^{-1}$ and at 283K $N_{\textup{tb}} = 4.0 \times 10^{7} \  \textup{e}^{-}\textup{s}^{-1}$. 
    \item The \textit{quantization noise}, $N_{\textup{quant}}$, is due to the finite number of bits used for analogue-to-digital encoding of the signal detected by the detector chip. $N_{\textup{quant}}$ is calculated as, 
    \begin{equation}
         N_{\textup{quant}} = \frac{S_{\textup{FWC}}(i)}{\sqrt{12}\cdot 2^{\textup{nbits}}\left ( \frac{\Delta V_{\textup{ADC,usable}}}{\Delta V_{\textup{ADC,max}}} \right )}.
    \end{equation}
    $S_{\textup{FWC}}(i)$ is the full well capacity of the detector pixel $i$. $\Delta V_{\textup{ADC,usable}}$ corresponds to the usable voltage range of the analogue-to-digital converter and $\Delta V_{\textup{ADC,max}}$ is the converters maximum voltage range. For NOMAD $S_{\textup{FWC}}(i) = 3.7 \times 10^{7}$, $\Delta V_{\textup{ADC,max}} = 5 \ \textup{V}$, $\Delta V_{\textup{ADC,usable}} = 3.78 \ \textup{V}$ and $\textup{nbits} = 14$. As all terms are defined, $N_{\textup{quant}}$ can be calculated as, 
    
    \begin{equation}
         N_{\textup{quant}} = \frac{3.7 \times 10^{7}}{\sqrt{12}\cdot 2^{\textup{14}}\left ( \frac{3.78}{5.0} \right )} = 862.3 \ \textup{e}^{-}\textup{pixel}^{-1} \ \textup{(1d.p.)}.
    \end{equation}
    
    \item The \textit{shot noise}, $N_{\textup{shot}}$, is the noise resulting from the signal photons impacting the detector. $N_{\textup{shot}}$ is given by, 
    \begin{equation}
        N_{\textup{shot}} =\sqrt{S_{\textup{electrons}}(i).}
    \end{equation}
    
     In the model applied, \cite{Tho:16}, flux is calculated one diffraction order at a time. Flux outside the gratings full spectral range are ignored as they are accounted for by the adjacent orders. The number of electrons $S(i)$ generated in a detector pixel during integration time $\Delta t$ is given by, 

    \begin{equation}
    S_{\textup{electrons}}(i) = \frac{\lambda _{i}}{hc}\tau _{\textup{det}}(\lambda _{i})E_{Q}(\lambda _{i})S(\lambda _{i})\Delta t
    \end{equation}.
    
    $\lambda _{i}$ is the wavelength corresponding to pixel $i$, $E_{Q}(\lambda _{i})$ is the quantum efficiency of the detector and $\tau _{\textup{det}}$ is the sensitivity of the detector of the given wavelength, including responsivity, cold filter transmission and detector window transmission.
    
    \begin{equation}
    S(\lambda ) = \frac{a_{p}\pi}{4(F/\#)}\Delta \lambda W_{\textup{slit}}\tau _{\textup{opt}}(\lambda )R(\lambda).
    \end{equation}
    
    $F/\#$ is the channels's F-number and $a_{p}$ is the area of one detector pixel, $\Delta \lambda$ is the pixel spectral bandwidth, $W_{\textup{slit}}$ is the slit width, $R(\lambda)$ is the incoming radiance, $\tau _{\textup{opt}}(\lambda )$ is transmission of the optics, which is the product of the transmission properties of all the optics, including the mirrors, lenses, AOTF $(\tau _{\textup{opt}}(AOTF)$), angular dependency and the echelle grating blaze function. 
    
    Although stated earlier, a clear distinction should be made between $S_{\textup{electrons,n}}$ and $S_{\textup{electrons}}(i)$, both are different \\ sources of electrons. $S(\lambda)$ and subsequently $S_{\textup{electrons}}$ are used to simulate the shot noise, $N_{\textup{shot}}$. 
    
\end{enumerate}

\subsubsection{Thermal emission model}

The thermal emission model employed in Ares follows that of \cite{Tho:16}. The total thermal background is difficult to calculate give that all components of NOMAD act as a blackbody radiation source, emitting thermal photons. Whilst no distinction is made in the thermal background estimation section of \cite{Tho:16} it is stated that all the components are assumed to be at the same temperature as the instrument. Optical components within NOMAD are attached to a thermally conductive baseplate away from the heat generating central electronics board. Because of this the temperature of the optical parts, should not deviate more than a few degrees from the main instrument body. The calculation of the thermal background estimation is split into three ordered sections and is run for each detector pixel individually in the following manner: 

\begin{enumerate}
    \item From the optical components before the AOTF.
    \item From the AOTF to the slit.
    \item From the slit to the detector.
\end{enumerate}

The thermal background contribution is modelled using the Planck function, and expressed as the power emitted per unit area per solid angle of NOMAD's component surface, at a given wavelength $\lambda$ and temperature $T$:

\begin{equation}
    B(\lambda,T) = \frac{2hc}{\lambda ^{5}}\left ( \frac{1}{\exp (\frac{hc}{\lambda k_{b}T}) - 1} \right). 
\end{equation}

The number of emission generated electrons per wavelength, $\lambda$, from surface $n$ is therefore given by, 
 
\begin{equation}
   S_{\textup{e}^{-},\textup{n}}(\lambda ,T_{n}) = \varepsilon _{\textup{n}}(\lambda ) B(\lambda, T_{\textup{n}})\left ( \frac{\lambda }{hc} \right )A\Omega \left ( \prod_{j=n+1}^{m} \tau_{j}(\lambda )  \right ) E_{R}(\lambda )\Delta t,
\end{equation}

\noindent where $\varepsilon _{\textup{n}}(\lambda )$ is the surface emissivity, $\left ( \frac{\lambda }{hc} \right )$ is the energy of a photon of wavelength $\lambda$. $A\Omega$ is the instrument entendue. $\prod_{j=n+1}^{m} \tau_{j}(\lambda )$ is the cumulative transmission of all $m$ optical elements between the surface and the pixel, $E_{Q}(\lambda)$ is the quantum efficiency of the detector and $\Delta t$ is the observation time. 

Firstly, the signal in electrons in the detector pixel $i$ by thermal emission before the AOTF can be calculated as follows, summed over the $n$ optical components, before the AOTF, and is given by, 

\begin{equation}
S_{\textup{tb1}}(i) = \sum_{n}^{ } \left ( \frac{S_{\textup{electrons,n}}(\lambda_{i+1},T_{n})+ S_{\textup{electrons,n}}(\lambda_{i},T_{\textup{n}})}{2} \right )  (\lambda _{i+1}-\lambda _{i}).
\end{equation}

Secondly, for components between the AOTF and the slit, the calculation becomes more complex, as all diffraction orders $x$ need to be accounted for. For components between the AOTF and the slit, the thermal background noise is given by, 

\begin{equation}
\begin{aligned}
    & S_{\textup{tb2}}(i) = \sum_{n}^{ } \sum_{x=0}^{\infty}\left ( \frac{S_{\textup{electrons,n}}(\lambda_{x,i+1},T_{\textup{n}}) S_{\textup{electrons,n}}(\lambda_{x,i},T_{\textup{n}}) }{2} \right ) \\ 
    & \times \ (\lambda_{x,i+1}-\lambda_{x,i}). \\
\end{aligned}
\end{equation}

Thirdly, for the grating and components after the slit the signal in electrons is given by, 

\begin{equation}
    S_{\textup{tb3}}(i) = \sum_{n}^{ }\int_{\lambda=0}^{\infty} S_{\textup{electrons,n}}(\lambda, T_{\textup{n}}).
\end{equation}

Therefore, the total number of electrons generated in detector pixel $i$ due to the thermal emission from NOMAD is given by, 

\begin{equation}
    S_{\textup{tb}}(i) = \sum_{j=1}^{3} S_{\textup{tbj}}(i).
\end{equation}

The optical design of the NOMAD SO channel is such that it is composed of 7 elements, namely, the entrance optics, the AOTF filter, the spectrometer entrance slit, the collimating / imaging parabolic mirror, the echelle grating, the folding mirror and the detector. Returning to the problem of simulating $N_{\textup{shot}}$ and $N_{\textup{tb}}$, we are faced with the challenge of defining all of the aforementioned terms. 

As stated earlier, the signal arriving on the detector can be expressed as, 

\begin{equation}
S(\lambda ) = \frac{a_{p}\pi}{4(F/\#)}\Delta \lambda W_{\textup{slit}}\tau _{\textup{opt}}(\lambda )R(\lambda).
\end{equation}

The term $R(\lambda)$ is the incoming radiance from the Sun at Mars distance. This term is simulated with the Ares forward model. $\tau _{\textup{opt}}$ is the transmission of the optics, and is a product of the optics, mirrors, lenses, AOTF ($\tau _{\textup{AOTF}}$), the angular dependency and the echelle grating blaze function, given by,

\begin{equation}
\tau _{\textup{opt}}(\lambda) = \prod_{j=n+1}^{m} \tau _{j}(\lambda).
\end{equation}

$F/\#$ is the channel's F-number, the F-number of an optical system is the ratio of the system's focal length to the diameter of the entrance aperture. The F-number is commonly indicated with the $f/N$ format where $N$ is the F-number. For NOMAD SO the limiting F-number is $f/5.12$, \cite{Nef:15}. Note that in Table 10 of \cite{Nef:15} this is given for the cold shield aperture as $f/3.936$. $a$ is the area of one detector pixel, which \cite{Nef:15} gives as $30 \times 30 \ \mu \textup{m}^{2}$. $\Delta \lambda$ is the pixel spectral bandwidth, which can be determined from Equation 1 of \cite{Liu:19}. The size of the slit in the NOMAD SO channel is $60 \ \mu \textup{m} \times 900 \ \mu \textup{m}$, therefore $W_{\textup{slit}}= 60 \ \mu \textup{m}$. 

\section{Glossary}

\begin{table*}[ht]
\textbf{Appendix B. Glossary}
\centering
\begin{tabular}{p{0.2\linewidth}p{0.5\linewidth}p{0.2\linewidth}}
\hline
Variable & Description & Example Equation(s) \\
\hline
$m$ & AOTF diffraction order number. & 1 \\
$p$ & Pixel number, $p\in [0,319]$. & 1 \\
$\nu$ & Wavenumber in $\textup{cm}^{-1}$ & 1 \\
$F_{0}$,$F_{1}$, $F_{2}$ & NOMAD SO $p,\nu,m$ relation coefficients. & 1 \\
$G_{0}$,$G_{1}$, $G_{2}$ & NOMAD SO tuning relation coefficients. & 1 \\
$A$ & AOTF frequency. & 2 \\
$TF$ & AOTF transfer function. & 3 \\
$\nu_{0}$ & AOTF transfer function centre in $\textup{cm}^{-1}$. & 3 \\
$w$ & First zero crossing of sinc-squared function. & 3 \\
$I_{G}$ & Gaussian amplitude of the AOTF transfer function. & 3 \\
$\rho_{G}$ & Gaussian standard deviation of the AOTF transfer function. & 3 \\
$q$,$n$ & AOTF transfer function continuum offset parameters. & 3 \\
$I_{0}$ & Sinc-squared amplitude of the AOTF transfer function. & 4 \\
$F_{\textup{sinc}}$ & AOTF transfer function sinc-squared contribution. & 4 \\
$F_{\textup{gauss}}$ & AOTF transfer function Gaussian contribution. & 5 \\
$F_{\textup{cntmn}}$ & AOTF transfer function continuum contribution. & 6 \\
$F_{\textup{blaze}}$ & Blaze function. & 7 \\
$p_{0}$ & Blaze function pixel centre in pixel units. & 7 \\
$w_{p}$ & Blaze function width. & 7 \\
$I_{0}$,$I_{1}$ & Blaze function $p_{0}$ relation coefficient. & 8 \\
$PEC$ & $\textit{Partial Elements Continuum}$. & 9 \\
$PE$ & A $\textit{Partial Element}$ of the $PEC$. & 9 \\
$\delta m$ & A discrete change in diffraction order $m$. & 9 \\
$AOTF$ & AOTF transfer function at the AOTF frequency $A$, for spectral grid $\nu_{j}$ of diffraction order $j$. & 10 \\
$R(A, \mathbf{\nu_{m}})$ & Observed Radiance by the NOMAD SO channel. & 11 \\
$R(j, \mathbf{\nu_{j}})$ & Signal terms. & 11 \\
$\mathbf{\nu_{j}}$ & Spectral grid of diffraction order $j$ & 10,11,12 \\
$n_{\textup{lat-lon}}$ & Ares Geometry module grid point number. & \\
$n_{\textup{layers}}$ & Number of atmospheric layers. & \\
$r_{\textup{s}}$ & TGO observation altitude. & 13,14,15\\
$\varphi_{\textup{s}}$ & TGO sub-observation longitude. & 13,14,15 \\
$\theta_{\textup{s}}$ & TGO sub-observation latitude. & 13,14,15 \\
($x_{\textup{s}}$,$y_{\textup{s}}$,$z_{\textup{s}}$) & TGO Cartesian coordinate. & 13,14,15 \\
($x_{\textup{t}}$,$y_{\textup{t}}$,$z_{\textup{t}}$) & Tangent point Cartesian coordinates. & \\
$\in$ & Set membership, read as \textit{is an element of}. & \\
$\forall$ & Universal quantifier, read as \textit{for all}. &  \\
$r_{\textup{Mars}}$ & Mars radius in km. &  18 \\
$\mathbf{y}$ & Measurement vector. & 22,28,29,30,31,32 \\
$\mathbf{x}$ & State vector. & 22,28,29,30,31,32 \\
$\mathbf{b}$ & Parameter vector. & 22 \\
$\mathbf{F(x,b)}$ & Forward model. & 22 \\
$\epsilon$ & Error vector in the forward model & 22 \\
$\lambda$ & Wavelength. & 23,24,25,26,29 \\
$k_{\textup{B}}$ & Boltzmann constant.  & A.10 \\
$\tau$ & Optical depth. &   \\
$P(\mathbf{x},\mathcal{M})$ & pdf of $\mathbf{x}$ for atmospheric forward model $\mathcal{M}$. & 28,30 \\
$P(\mathbf{y},\mathcal{M})$ & pdf of $\mathbf{y}$ for atmospheric forward model $\mathcal{M}$. & 28,30 \\
$P(\mathbf{y}|\mathbf{x},\mathcal{M})$ & pdf of $\mathbf{y}$ given $\mathbf{x}$ for atmospheric forward model $\mathcal{M}$. & 28 \\
$P(\mathbf{x}|\mathbf{y},\mathcal{M})$ & $\textit{a posteriori}$ pdf of $\mathbf{x}$ given $\mathbf{y}$ for atmospheric forward model $\mathcal{M}$. & 29 \\
$\mathbf{0}$ & Zero vector. & 31 \\
$\nabla_{\mathbf{x}}$ & Gradient operator for $\mathbf{x}$. & 31 \\
$l(z)$ & Optical path length. & 24 \\
$N_{m}$ & Total number of absorbing species. & 25 \\
$\mathcal{T}_{\lambda}$ & Monochromatic transmittance at wavelength $\lambda$. & 26 \\
\hline
\end{tabular}
\end{table*}

\begin{table*}[ht]
\textbf{Appendix B. Glossary (Continued)}
\centering
\begin{tabular}{p{0.2\linewidth}p{0.5\linewidth}p{0.2\linewidth}}
\hline
Variable & Description & Example Equation(s) \\
\hline
$dl_{j}$ & Atmospheric path lengths. & \\
$a$,$b$,$c$ & Ellipsoid coefficients. & 17 \\
$r_{\textup{Mars,p}}$ & Mars polar radius in km. & 18 \\
$r_{\textup{Mars,e}}$ & Mars equatorial radius in km. & 18 \\
$f$ & Ellipsoid flattening ratio. & 18 \\
$\Delta z_{i}$ & Height of atmospheric layer $i$ above the surface ellipsoid. & 19,20,21 \\
$N$ & NOMAD total noise contributions. & A.1 \\
$N_{\textup{dark}}$ & Dark noise. & A.1,A.2 \\ 
$N_{\textup{shot}}$ & Shot noise. & A.1,A.7 \\
$N_{\textup{ro}}$ & Readout noise. & A.1,A.3 \\
$N_{\textup{quant}}$ & Quantisation noise. & A.1,A.5,A.6 \\ 
$N_{\textup{tb}}$ & Thermal background noise. & A.1 \\ 
$\Delta t$ & Integration time in seconds. & A.2 \\
$S_{\textup{tb}}(i)$ & Total number of electrons generated in detector pixel $i$ due to instrument thermal emission. & A.4 \\
$S_{\textup{FWC}}(i)$ & Detector pixel $i$ full-well-capacity. & A.5 \\
$\Delta V_{\textup{ADC,usable}}$ & Usable voltage range of the analogue-to-digital converter. & A.5 \\
$\Delta V_{\textup{ADC,max}}$ & Maximum voltage range of the analogue-to-digital converter. & A.5 \\
$S_{\textup{electrons}}(i)=S(i)$ & Number of electrons generated in detector pixel $i$ during integration time, associate with calculating  $N_{\textup{shot}}$. & A.7 \\
$\lambda_{i}$ & Wavelength corresponding to pixel $i$. & A.8 \\
$E_{Q}(\lambda_{i})$ & Detector quantum efficiency at $lambda_{i}$. & A.8 \\
$\tau_{\textup{det}}(\lambda_{i})$ & Detector sensitivity at wavelength $\lambda_{i}$. & A.8 \\
$F/\#$ & Channel F-number. & A.9 \\
$a_{p}$ & Area of one detector pixel. & \\
$\Delta \lambda$ & Pixel spectral bandwidth. & \\
$W_{\textup{slit}}$ & Detector slit width. & A.9 \\
$R(\lambda)$ & Incoming radiance. & \\
$\tau_{\textup{opt}}(\lambda)$ & Detector optics transmission. & A.9 \\
$\tau_{\textup{opt}}(AOTF)$ & AOTF optics transmission. & \\
$S_{\textup{electrons,n}}$ = $S_{\textup{e}^{-},\textup{n}}$ & The number of emission generated electrons per wavelength $
\lambda$ from surface $n$. & A.11 \\
$\prod_{j=n+1}^{m} \tau_{j}(\lambda )$ & Cumulative transmission of all $m$ optical elements between the surface and the pixel. & A.11  \\ 
$E_{Q}(\lambda)$ & Quantum efficiency of the detector. & A.11 \\ 
$\Delta t$ & Observation time. & A.11 \\
$S_{\textup{tb1}}(i)$ & Detector pixel $i$ thermal emission signal in electrons before the AOTF. & A.12 \\ 
$S_{\textup{tb2}}(i)$ & Detector pixel $i$ thermal emission signal in electrons between AOTF and slit. & A.13 \\
$S_{\textup{tb3}}(i)$ & Detector pixel $i$ thermal emission signal in electrons between slit, grating and components. & A.14 \\
$I_{\lambda}(z) $ & The monochromatic intensity of radiation passing through a gas. & \\
$I_{\lambda}(0) $ & The monochromatic intensity of radiation at the top of the atmosphere. & \\
$\tau_{\lambda}(z) $ & Medium optical depth at altitude $z$. & \\
$\varepsilon _{\textup{n}}(\lambda )$ & & \\
$A\Omega $ & NOMAD instrument entendue. & A.11 \\ 
$T$ & Temperature in Kelvin (K). & \\
$h$ & Planck constant. & A.8,A.10,A.11 \\
$c$ & Speed of light in a vacuum. & A.8,A.10,A.11 \\
$B(\lambda, T)$ & Planck function. & A.10 \\
$\zeta_{m}(\lambda)$ & Absorption cross-section at wavelength $\lambda$. & 24 \\
$\chi_{m}(z)$ & Atmospheric column density for molecule $m$. & 24 \\
$\tau _{\lambda }(z)$ & Total optical depth. 25 & \\
\hline
\end{tabular}
\end{table*}

\section{Abbreviations}

\begin{table*}[ht]
\textbf{Appendix C. Abbreviations}
\centering
\begin{tabular}{p{0.45\linewidth}p{0.45\linewidth}}
\hline
Variable & Description \\
\hline
\textbf{NOMAD} & \textbf{N}adir and \textbf{O}ccultation for \textbf{MA}ars \textbf{D}iscovery \\
\textbf{TGO} & \textbf{T}race \textbf{G}as \textbf{O}rbiter \\
\textbf{ESA} & \textbf{E}uropean \textbf{S}pace \textbf{A}gency \\
\textbf{ppmv} & \textbf{p}arts \textbf{p}er \textbf{m}illion per unit \textbf{v}olume \\
\textbf{ppbv} & \textbf{p}arts \textbf{p}er \textbf{b}illion per unit \textbf{v}olume \\
\textbf{pptv} & \textbf{p}arts \textbf{p}er \textbf{t}rillion per unit \textbf{v}olume \\
\textbf{RTM} & \textbf{R}adiative \textbf{T}ransfer \textbf{M}odel \\
\textbf{MSL} & \textbf{M}ars \textbf{S}cience \textbf{L}aboratory \\
\textbf{SAM} & \textbf{S}ample \textbf{A}nalysis at \textbf{M}ars \\
\textbf{TLS} & \textbf{T}unable \textbf{L}aser \textbf{S}pectrometer \\
\textbf{FTS} & \textbf{F}ourier \textbf{T}ransform \textbf{S}pectrometer \\
\textbf{RTE} & \textbf{R}aditive \textbf{T}ransfer \textbf{E}quation \\
\textbf{MAP} & \textbf{M}aximum \textbf{A} \textbf{P}osteriori \\
\textbf{MCD} & \textbf{M}ars \textbf{C}limate \textbf{D}atabase \\
\textbf{PSA} & \textbf{P}lanetary \textbf{S}cience \textbf{A}rchive \\
\textbf{GCM} & \textbf{G}lobal \textbf{C}limate \textbf{M}odel \\
\textbf{NASA} & \textbf{N}ational \textbf{A}eronautics and \textbf{S}pace \textbf{A}dministration \\
\textbf{BIRA-IASB} & The Royal Belgian Institute for Space Aeronomy \\
\textbf{ILS} & \textbf{I}nstrument \textbf{L}ine \textbf{S}hape \\
\textbf{HDF} & \textbf{H}ierarchical \textbf{D}ata \textbf{F}ormat \\
\textbf{SNR} & \textbf{S}ignal to \textbf{N}oise \textbf{R}atio \\
\textbf{XML} & \textbf{X}tensible \textbf{M}arkup \textbf{L}anguage \\
\textbf{AOTF} & \textbf{A}cousto \textbf{O}ptic \textbf{T}unable \textbf{F}ilter \\
\textbf{EAICD} & \textbf{E}xperiment to \textbf{A}rchive \textbf{I}nterface \textbf{D}ocument \\
\textbf{SO} & \textbf{S}olar \textbf{O}ccultation \\
\textbf{LNO} & \textbf{L}imb \textbf{N}adir \textbf{O}ccultation \\
\textbf{PFS} & \textbf{PF}lanetary \textbf{F}ourier \textbf{S}pectrometer \\
\textbf{MCMC} & \textbf{M}arkov \textbf{C}hain  \textbf{M}onte \textbf{C}arlo  \\
\textbf{LOS} & \textbf{L}ine \textbf{O}f  \textbf{S}ight \\
\textbf{VMR} & \textbf{V}olume \textbf{M}ixing \textbf{R}atio \\
\textbf{UVIS} & \textbf{U}ltra \textbf{V}iolet \textbf{I}nfrared \textbf{S}pectrometer \\
\textbf{FWHM} & \textbf{F}ull \textbf{W}idth \textbf{H}alf \textbf{M}aximum \\
\textbf{PSG} & \textbf{P}lanetary \textbf{S}pectrum \textbf{G}enerator \\
\textbf{MGS} & \textbf{M}ars \textbf{G}lobal \textbf{S}urveyor \\
\textbf{MOLA} & \textbf{M}ars \textbf{O}rbiter \textbf{L}aser \textbf{A}ltimeter \\
\textbf{DSK} & \textbf{D}igital \textbf{S}hape \textbf{K}ernel \\
\textbf{API} & \textbf{A}pplication \textbf{P}rogramming \textbf{I}nterface \\
\textbf{HAPI} & \textbf{H}itran \textbf{A}pplication \textbf{P}rogramming \textbf{I}nterface \\
\textbf{NEMESIS} &  \textbf{N}on-linear Optimal \textbf{E}stimator for \textbf{M}ultivariat\textbf{E} \textbf{S}pectral. \textbf{A}naly\textbf{SIS}.\\
\hline
\end{tabular}
\end{table*}

\section{Chemical Species}

\begin{table*}[ht]
\textbf{Appendix D. Chemical Species}
\centering
\begin{tabular}{p{0.45\linewidth}p{0.45\linewidth}}
\hline
Symbol & Description \\
\hline
HCN & Hydrogen Cyanide \\
OCS & Carbonyl Sulfide \\
$\textup{S}\textup{O}_{2}$ & Sulfur Dioxide \\
$\textup{N}\textup{H}_{3}$ & Ammonia \\
$\textup{H}\textup{O}_{2}$ & Hydroperoxyl \\
$\textup{HCl}$  & Hydrogen Chloride  \\
$\textup{H}_{2}\textup{S}$ & Hydrogen Sulfide \\
$\textup{H}_{2}\textup{CO}$ & Formaldehyde \\
$\textup{N}_{2}\textup{O}$  & Nitrous Oxide \\
$\textup{N}\textup{O}_{2}$  & Nitrogen Dioxide \\
$\textup{C}_{2}\textup{H}_{4}$  & Ethylene (Ethene) \\
$\textup{C}_{2}\textup{H}_{2}$ & Acetylene (Ethyne) \\
HDO & Deuterium Oxide (Heavy water) \\
$\textup{CH}_{4}$ & Methane \\
$\textup{C}_{2}\textup{H}_{6}$ & Ethane \\
$\textup{C}_{3}\textup{H}_{8}$ & Propane \\
$\delta ^{13}\textup{C}$ & Delta Carbon Thirteen \\
$\delta ^{2}\textup{H}$ & D/H ratio \\
$\textup{CO}$ & Carbon Monoxide \\
$\textup{CO}_{2}$ & Carbon Dioxide \\
$\textup{H}_{2}\textup{O}$ & Water \\
$\textup{O}_{3}$ & Ozone \\
\hline
\end{tabular}
\end{table*}

\end{document}